\documentclass[pra,twocolumn,showpacs,nofootinbib]{revtex4}
\usepackage{amssymb,amsmath}
\usepackage{epsfig}
\usepackage{graphicx}
\usepackage{rotating}  
\def\shiftdown#1{#1\llap{\lower.04ex\hbox{#1}}}

\begin{document}
\title{Collision-induced radiative quenching and other disintegration modes 
of the $2s$ state of muonic hydrogen and deuterium atoms} 
\author{V.P. Popov}
\email{vlad.popov1945@mail.ru}
\author{V.N. Pomerantsev}
\email{pomeran@nucl-th.sinp.msu.ru}

\affiliation{Skobeltsyn Institute of Nuclear Physics,
 Lomonosov 
Moscow State University, 
119991 Moscow, Russia}           

\date{\today}    

\begin{abstract}
The formation and various disintegration modes of  $2s$ states for muonic hydrogen and
deuterium atoms at kinetic energies both above and below the $2s-2p$ threshold are
studied. The cross sections of the collision-induced radiative quenching, elastic
scattering, and Coulomb deexcitation of $(\mu^- p)_{2s}$ and $(\mu^- d)_{2s}$ atoms in
collisions with ordinary hydrogen and deuterium atoms at collision energies below the
$2s-2p$ threshold are calculated in the framework of the close-coupling approach.  
The basis set includes all open and closed channels corresponding to exotic-atom
states with principal quantum number $n$ from 1 up to 30.   Results of these numerical
quantum-mechanical calculations of cross sections are used as input data for detailed
cascade calculations. The kinetics of atomic cascade for $\mu^- p$ and  $\mu^-d$ atoms
is studied in the wide range of the relative target densities, $\varphi = 10^{-8} -1$
applying the improved version of the cascade model.   It is shown that the
collision-induced radiative quenching gives a significant contribution to the total
disintegration of the $2s$ state of muonic atoms at energies below the  $2s-2p$
threshold.  The initial population of the $2s$ state for both muonic hydrogen and
deuterium atoms, absolute intensities and probabilities of the different
disintegration modes as well as their lifetimes are calculated for the fractions with
kinetic energies above and below the $2s-2p$ threshold. The obtained results are
compared with the known experimental data.
\end{abstract}
\maketitle

\section {Introduction}
The negatively charged muons stopped in $H_2$ or $D_2$ gaseous target form
$\mu^- p$ or $\mu^- d$ atoms in highly excited states \cite{KP, KPF,JC}. The
formation of muonic atoms of hydrogen isotopes is followed by a number of the
subsequent radiative and collisional processes.  In the process of deexcitation,
an $\mu^-p$ or $\mu^-d$  atom passes through many intermediate states until it
reaches the ground state or weak decay of the meson occurs ($\mu^{-} \to e^- +
\bar{\nu_e} + \nu_{\ mu}$) in the excited state of the atom. During this
deexcitation cascade, a fraction of  $\mu^-  p$ or $\mu^- d$ atoms reaches the
$2s$ state.  The number of atoms reached $2s$ state is determined by several factors: the
initial $(n,\,l,\,E)$-distributions (e.g., see \cite{PPForm, PPIsot}), the
competition of collisional processes and radiative transitions during the
deexcitation cascade, as well as the target temperature and density.

 Muonic hydrogen and  muonic deuterium atoms are of special interest among
exotic atoms due to  their simplest structure and possibility to investigate a
number of problems  both the exotic atom physics and the bound-state QED. The
$2s$ state plays a particular role in these atoms due to $(2s-2p)$  Lamb shift, 
$E_L=202.0$ meV, and has no analog in hadronic atoms.  A high-energy component
of $(\mu p)_{1s}$ with kinetic energy  $\sim 0.9$ keV was discovered
\cite{RPD,PPK}  from the  analysis of the time-of-flight spectra (at low gas
pressures $p_{H_2}$ = 16 and 64 hPa) and attributed  to non-radiative quenching
of the $2s$ state due to the formation of the muonic molecule in a resonant
$(\mu p)_{2s}+{\rm H}_2$ collision  and subsequent Coulomb deexcitation of the
$(pp\mu)^+$ complex  (see \cite{WJKF} and references therein). 
Theoretical estimation of the quenching rate in the framework of this model 
gives $\sim 5 \times 10^{10}$ $s^{-1}$ at liquid-hydrogen  density (LHD=4.25
$\times 10^{22}$ atoms/cm$^{3}$) that is about  an order of magnitude smaller 
than the  value $4.4^{+2.1}_{-1.8} \times 10^{11} s^{-1}$ extracted from 
experiments \cite{PPK}. In our paper~\cite{PPForm} the observed collisional
quenching  and high-energy  $(\mu p)_{1s}$ component was explained by the direct
Coulomb deexcitation (CD) process,
\begin{equation} \label{eq. 1}
(\mu^- p)_{2s} + {\rm H} \to (\mu^- p)_{1s}{\rm( 0.9 keV)}  + {\rm H},
\end{equation}

The problem of the $(\mu^- p)_{2s}$ collisional quenching in the 
hydrogen target has been studied previously both 
experimentally \cite{Eg81, HA2, Breg, FKott, RPohl, RPD} and
theoretically \cite{KL71, Muel75, CF77, CohB81,CascalII, PPForm}.
Experimental investigations with muonic hydrogen
and deuterium atoms at Paul Scherrer 
Institute \cite{PDH, Ludh, Natur, Mud} performed
at a low gas pressure have revived interest in the subject of the 
formation and disintegration modes
of $(\mu^- p)_{2s}$ and $(\mu^- d)_{2s}$ atoms at kinetic energies
both above and below the $2s-2p$ threshold.

The theoretical description of the collisional processes and kinetics of 
the atomic cascade was improved significantly in our 
works (e.g., see \cite{PPForm, PPKin, KPP, diff, pion, PPInHyp, PPInd, PPIsot} and refs. therein). 

In the collision of muonic hydrogen or deuterium atoms in the $2s$ state with
the target atom or molecule, a strong coupling between this state and the
unstable with respect to the radiative $2p\to 1s$ transition $2p$ state leads to
the collision-induced radiative quenching of the $2s$ state at kinetic energies
below the $2s-2p$ threshold.  The consequences of this mechanism can be observed in
both the absolute x-ray yield of the $K\alpha$ line and the lifetime of the
muonic hydrogen and deuterium atoms in the $2s$ state. 
According to our
knowledge, the process of the collision-induced radiative quenching of the $(\mu
p)_{2s}$  state has been treated in the framework of the semiclassical
approaches \cite{Muel75, CohB81} and was never taken into account in cascade
calculations, since the delayed $K_{\alpha}$ x-rays induced during the
collisions have not been experimentally
observed \cite{HA2}.

In this paper, we investigate the further history 
of $(\mu^-p)$ and $(\mu^-d)$ atoms formed in the
$2s$ state with kinetic energies both above and below 
the $2s-2p$ threshold. Their fate is determined
by the interplay of a few processes: the elastic $2s-2s$ scattering,
the $2s\to2p$ Stark transition followed by the fast radiative 
$2p\to1s$ deexcitation, the collision-induced radiative quenching,
the Coulomb deexcitation $2s\to1s$, and $\mu$-decay.

The paper is organized as follows. Section II presents the results of calculations of the cross sections for elastic $ 2s-2s $ scattering, $ 2s-2p $ Stark transitions, collision-induced radiation quenching, and $ 2s-1s $ Coulomb deexcitation for muonic hydrogen and deuterium atoms performed in the  framework of the close-coupling approach.
Results of cascade calculations are presented in Sec. III. Initial populations and kinetic energy distributions of $(\mu^- p)_{2s}$ and $(\mu^- d)_ {2s}$ atoms are presented in Sec. III A. Collisional rates for $(\mu^- p)_{2s}$ and $(\mu^- d)_{2s}$ atoms and intensities of the different modes of their disintegration are given in Secs. III B and III C, correspondingly.  The contributions of different processes, forming the absolute x-ray yield of $K\alpha$ line are discussed in Sec. III D. Calculated probabilities of different disintegration modes and a lifetime of $(\mu^- p)_{2s}$ and $(\mu^- d)_{2s}$ atoms with energies both above and below the $2s-2p$ threshold are presented in Sec. III E. The results are summarized in Sec. IV.

Atomic units are used throughout unless otherwise stated.
  
\section {Cross sections of collisional processes for muonic hydrogen and deuterium atoms}

 In our paper \cite{PPIsot}, the cross
sections of all processes at collisions of the excited $\mu^- p$ and
$\mu^- d$ atoms with hydrogen isotope  atoms in the ground state have
been presented. In these calculations, performed in the  framework of the
close-coupling approach, the radiative widths or instability 
of the excited states of muonic atoms and the possibility of the 
collision-induced radiative transitions were not taken into account. 
This approximation is completely justified for all excited $nl$ states 
except the $2p$ state. Indeed, the radiative 
widths of highly excited states are very small and the rates 
of the collision-induced radiative
transitions are negligible compared to the rate of the direct radiative dipole
transitions $nl\to n'l\pm 1$.
 
The situation is quite different in the case of the $2s$ state. 
At collision energies below the $2s-2p$
threshold, the Stark transition $2s\to 2p$ is energetically forbidden and 
the collision-induced radiative quenching 
\begin{equation}
(\mu^{-} a)_{2s}+ A \to (\mu^{-} a)_{1s} +A + \gamma (1.9\, \rm {keV})
\label{indrad}
\end{equation}
($a=p$ or $a=d$, and $A=H, D$)
may prove to be an important mechanism of the disintegration of the
$2s$ state. Therefore, in this paper devoted to the fate 
of the $2s$ state in muonic hydrogen and deuterium atoms, 
the cross sections of the elastic scattering
\begin{equation}
(\mu^{-} a)_{2s}+ A \to (\mu^{-} a)_{2s} +A  
\label{elsc}
\end{equation}
 and Coulomb deexcitation (CD)
 \begin{equation}
(\mu^{-} a)_{2s}+ A \to (\mu^{-} a)_{1s}(0.9\, \rm {keV}) +A(1.0\, \rm {keV}) 
\label{cd}
\end{equation}
 in $(\mu^{-} p)_{2s}+ H$  and 
$(\mu^- d)_{2s} + D$ collisions calculated earlier\,\cite{PPForm,PPIsot}
at energies below the $2s-2p$ threshold were recalculated 
taking into account the radiation width or the instability of the $2p$ state.

Calculations were performed in the framework of the closed coupling approach by using the propagation 
matrix method described in details in our paper \cite{PPForm}. The main advantage of 
this method is that it allows to describe both open and closed  
unstable-state channels as well as to avoid numerical errors caused by
the exponentially increasing linearly-independent 
solutions of the system of the close-coupling equations. 
This method has been developed and 
applied earlier by the authors, in particular, for the description of the collision-induced absorption or annihilation 
in the case of hadronic atoms~\cite{PPInd, PPInHyp}.

The present calculations were performed with the extended basis set 
including all states of the muonic hydrogen (deuterium) atom
with the principal quantum number values  from $n=1$ 
up to $n _{max}= 30$. The value $\Gamma_{2p}=80$~$\mu$eV 
of the radiative width of the $2p$ state was used for both muonic 
hydrogen and muonic deuterium atoms. All other states 
were considered as stable ones.
In the case of the $2s$ state at kinetic energies below 
the $2s-2p$ threshold, the $S$ matrix in the subspace 
of the open channels is not unitary 
due to the instability of the $2p$ state. The unitary defect allows  
to determine the cross section of the collision-induced 
radiative transition from the $2s$ state as follows:
\begin{equation}  
\sigma^{\rm ind}_{2s}=\frac{\pi}{k^2}\sum_{J,n'=1,2}(2J+1)
\left(\delta_{2sJ,n'sJ}-\left|S^J_{2s\to n's}\right|^2\right). 
\label{sigmaind}  
\end{equation} 
 Here, $k^{2}=2M_r E_{\rm cm}$, $E_{\rm cm}$ is the relative motion 
energy in the entrance channel, and $M_r$ is the reduced mass
of the scattering problem. 

\begin{figure}[!ht]
\centerline{\includegraphics[width=0.47\textwidth,keepaspectratio]
{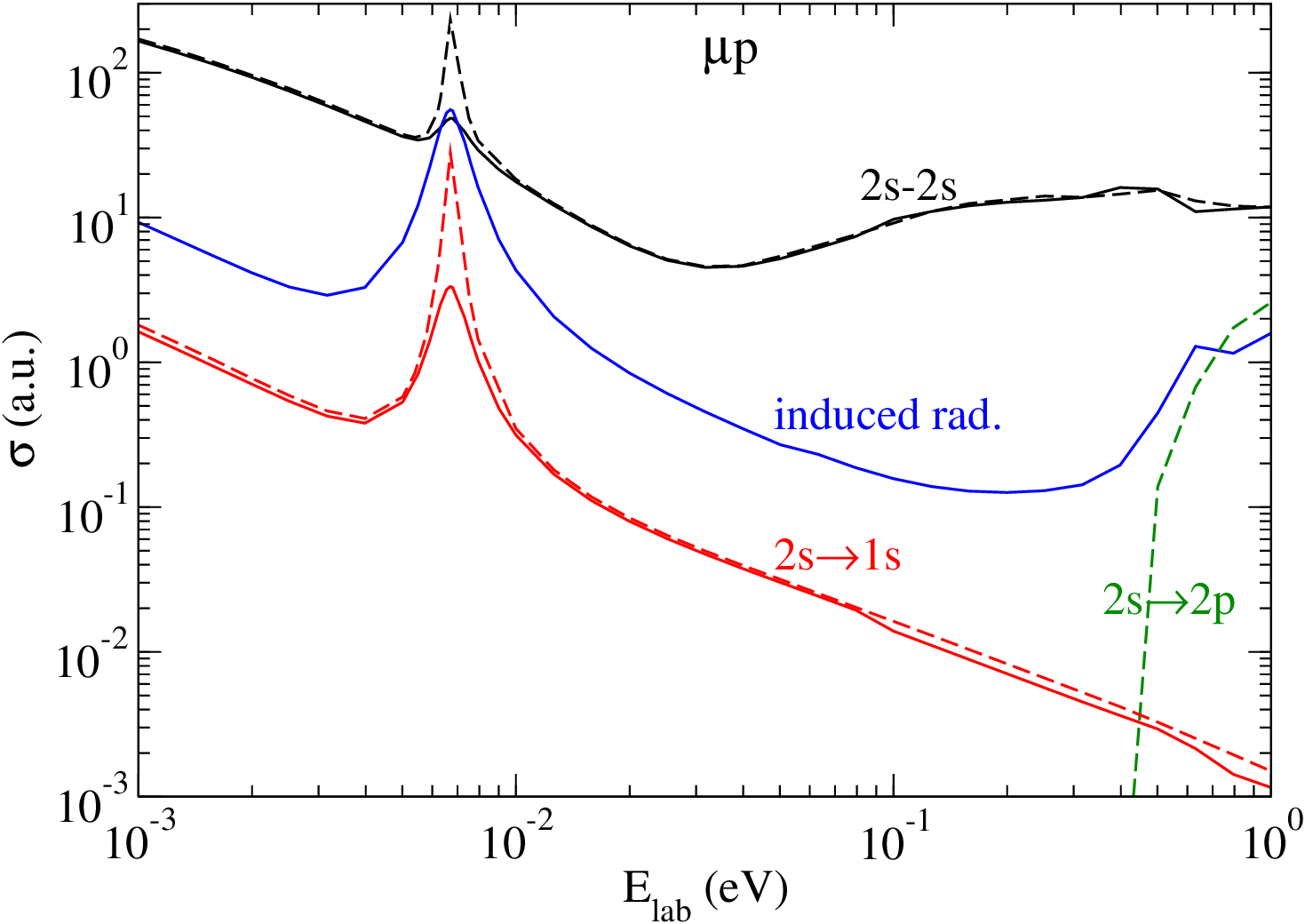}}
 \caption{  The energy dependence of
 cross sections for $(\mu^- p)_{2s} +H$ collisions:
 elastic scattering $2s - 2s$, 
 Coulomb deexcitation $2s\to1s$, collision-induced
 radiative quenching, and Stark $2s\to2p$ transition. 
 Cross sections of the elastic scattering 
 and Coulomb deexcitation calculated with 
 $\Gamma_{2p}=0$ \cite{PPForm} are shown by dashed lines for comparison.}
 \label{crsec_mup}
        \end{figure}

Cross sections for $(\mu^- p)_{2s} +H$ collisions calculated with
 $\Gamma_{2p}=80$~$\mu$eV and $\Gamma_{2p}=0$ are shown
 in Fig.~\ref {crsec_mup}.
As can be seen from Fig.\ref {crsec_mup}, the cross sections of the 
elastic scattering and CD practically do not change 
when the radiative width $\Gamma_{2p}$ of the $2p$ state is taken into account 
except for the resonance region. This means that the 
width can be considered as a small perturbation, and the cross section 
of the collision-induced process is proportional to the width. 
The dependence of the cross section of the collision-induced process 
on the value $\Gamma$ has been studied in detail in our 
paper \cite{PPInd}. At the same time, in the resonance region, 
the cross sections of the elastic scattering and the 
CD are strongly suppressed (by several times 
in comparison with results presented in \cite{PPForm}) due to 
the process (\ref{indrad}) of the collision-induced radiative quenching. In the 
resonance region, the cross section of the process (\ref{indrad})  
even exceeds the cross section of the elastic scattering.	
         
\begin{figure}[!ht]
\centerline{\includegraphics[width=0.47\textwidth,keepaspectratio]
{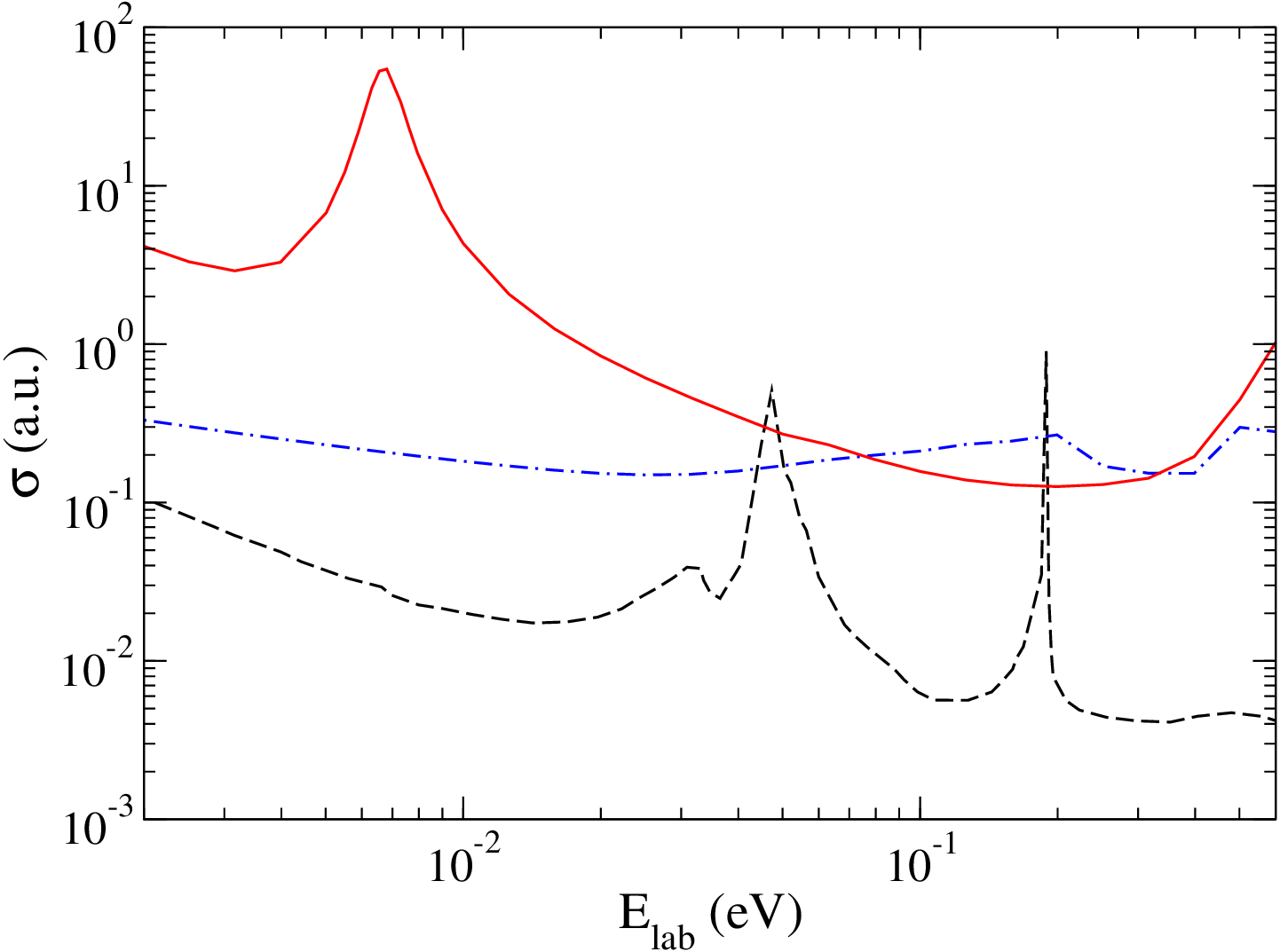}}
 \caption{  Cross sections of the collision-induced 
 radiative quenching for $(\mu^- p)_{2s} +H$ collisions vs 
 the laboratory kinetic energy below the $2s-2p$ 
 threshold: the solid line and dashed-dotted line are the results
 of the present calculations with the basis sets $n_{max} = 30$
 and $n_{max}=2$, respectively, and the dashed line is
 obtained from Fig.~4 of \cite{CohB81}.}
 \label{crsec_mup_comp}
        \end{figure}        
The comparison of the collision-induced radiative cross sections calculated in
the present fully quantum-mechanical approach with the 
cross sections calculated in \cite{CohB81}  within a simple model of the
potential scattering with the complex potential is shown in
Fig.\ref{crsec_mup_comp}.  Here the cross section of the collision-induced
radiative quenching calculated with the simplest basis set in which only wave
functions of the $1s$, $2s$, and $2p$ states of the muonic atom were included is
also presented.        In the paper \cite{CohB81}, the Born-Oppenheimer
approximation was applied using the four-term basis set with the lowest $1s$ and
$2s$ orbitals in variational calculations. In such approximation, the scattering
problem is reduced to a simple model of the potential scattering with the
complex potential. 

Figure 2 shows that the cross section of the collision-induced radiative
quenching calculated in  \cite{CohB81}  is too small even in comparison with our
close-coupling result obtained with the minimal basis set in which only the
$1s$, $2s$, and $2p$ muonic-atom wave functions were taken into account.
Besides, the cross section of collision quenching calculated in \cite{Muel75} is
somewhat less in comparison with the result of \cite{CohB81}. Therefore, the
adiabatic description and correspondingly the Born-Oppenheimer approximation is
not justified for the realistic treatment of the collision-induced radiative
quenching of the $2s$ state in $(\mu^- p)_{2s} +H$ collisions at kinetic
energies below the $2s-2p$ threshold. 

\begin{figure}[!ht]
\centerline{\includegraphics[width=0.47\textwidth,keepaspectratio]
{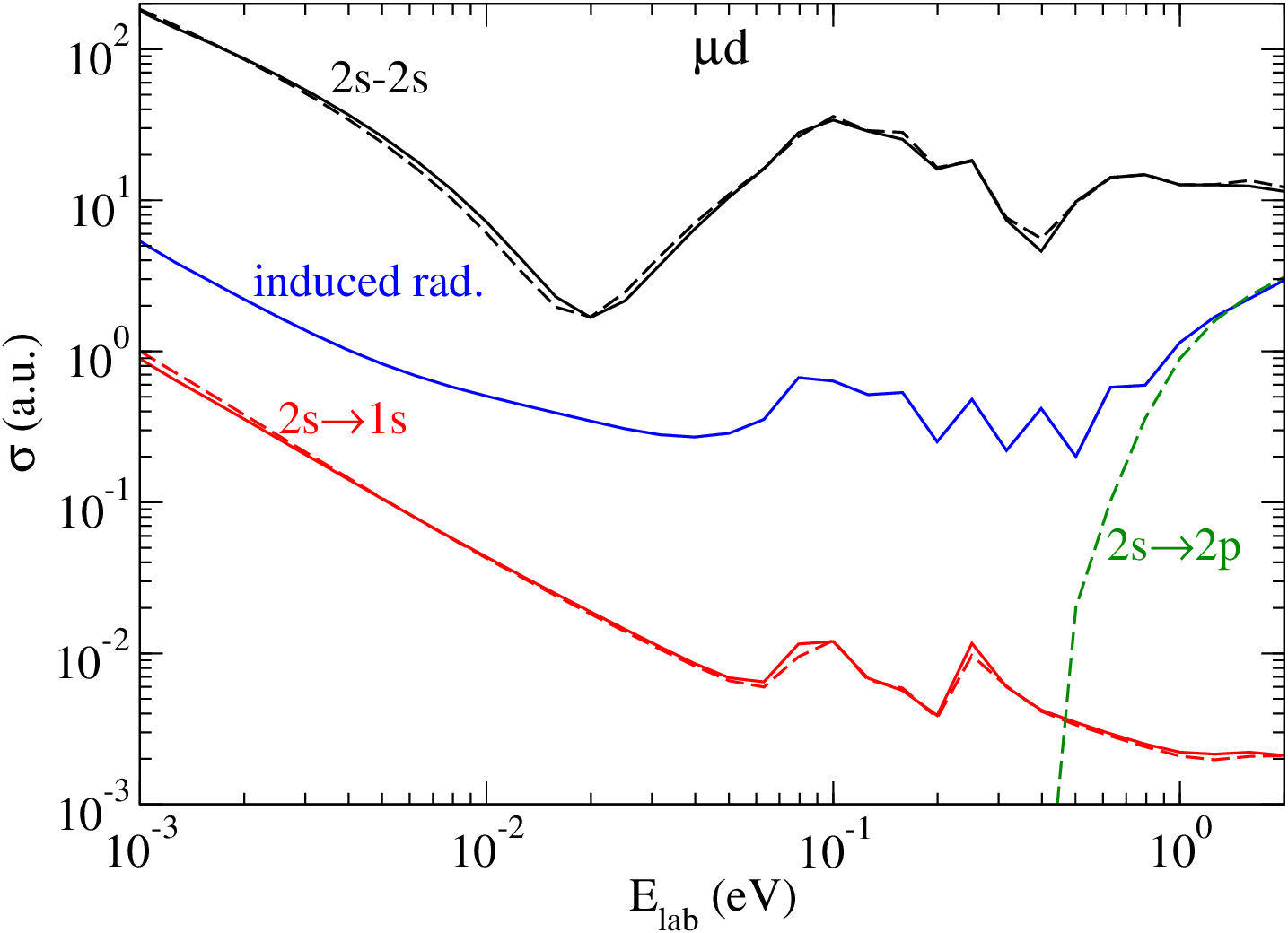}}
 \caption{  The same as in Fig.~\ref{crsec_mup}
 for $(\mu^- d)_{2s} +D$ collisions. Cross sections of the elastic scattering and Coulomb 
 deexcitation calculated with 
 $\Gamma_{2p}=0$ from~\cite{PPIsot} are shown by dashed lines.}
 \label{crsec_mud_2s}
        \end{figure}

Figure \ref{crsec_mud_2s} shows the energy dependence of the cross sections for
$(\mu^- d)_{2s} +D$ collisions. In contrast to the case of  the $(\mu^-
p)_{2s}$, there are no pronounced sub-threshold resonances, and, therefore,
taking into account the 2p-state width does not at all affect the elastic
scattering and CD cross sections. 
         
\section{Results of cascade calculations}

The cascade calculations for muonic hydrogen and muonic deuterium atoms were
performed in the framework of the improved version of the cascade model
developed in our recent papers\,\cite{PPForm, PPIsot}.  In particular, the
probability densities of $n$ and $l$  initial (at the instant of the exotic atom
formation) distributions as well as of the initial kinetic energy distribution
(see Eqs. 22-24 in \cite{PPIsot}) were used in the present cascade
calculations.  Besides, the exact kinematic of the binary collision taking into
account the target motion was applied in the present cascade calculations (for
detail see \cite{PPIsot}). Finally, the present cascade model includes new
results given in Sec.~II for cross sections of the elastic scattering, collision-induced
radiative quenching, and the CD for $(\mu^- p)_{2s}$ and $(\mu^- d)_{2s}$ at
kinetic energies below the $2s-2p$ threshold.

To obtain good statistics, the fate of the $10^8$ muonic atoms has been analyzed
in the present cascade calculations for each value of the target density in the
relative density range, covering nine orders of magnitude $\varphi = 10^{-9}
-1$, where $\varphi$ is the target density in units of the liquid hydrogen
density, $N_{lhd}=4.25\cdot 10^{22}$ atoms/cm$^3$. 

The various characteristics of muonic hydrogen and deuterium atoms in the $2s$
state were calculated: the arrival population and kinetic energy distribution,
the intensities and probabilities of the different disintegration modes, and
lifetimes. Some of present results are compared with the available experimental
data~\cite{HA2, PDH, RPD, Mud}.   
 
 \subsection {\bf Initial population and kinetic energy distribution 
 of $(\mu^- p)_{2s}$ and $(\mu^- d)_ {2s}$} 
  
We define the total initial or arrival population of the $2s$ state, $\varepsilon ^{\rm tot}_{2s}$, as the probability that the formed muonic hydrogen or deuterium atom will reach this state during the deexcitation cascade regardless of whether its kinetic energy is above or below of the $2s - 2p$ threshold. 

\begin{figure}[!ht]          
\centerline{
\includegraphics[width=\columnwidth,keepaspectratio]
{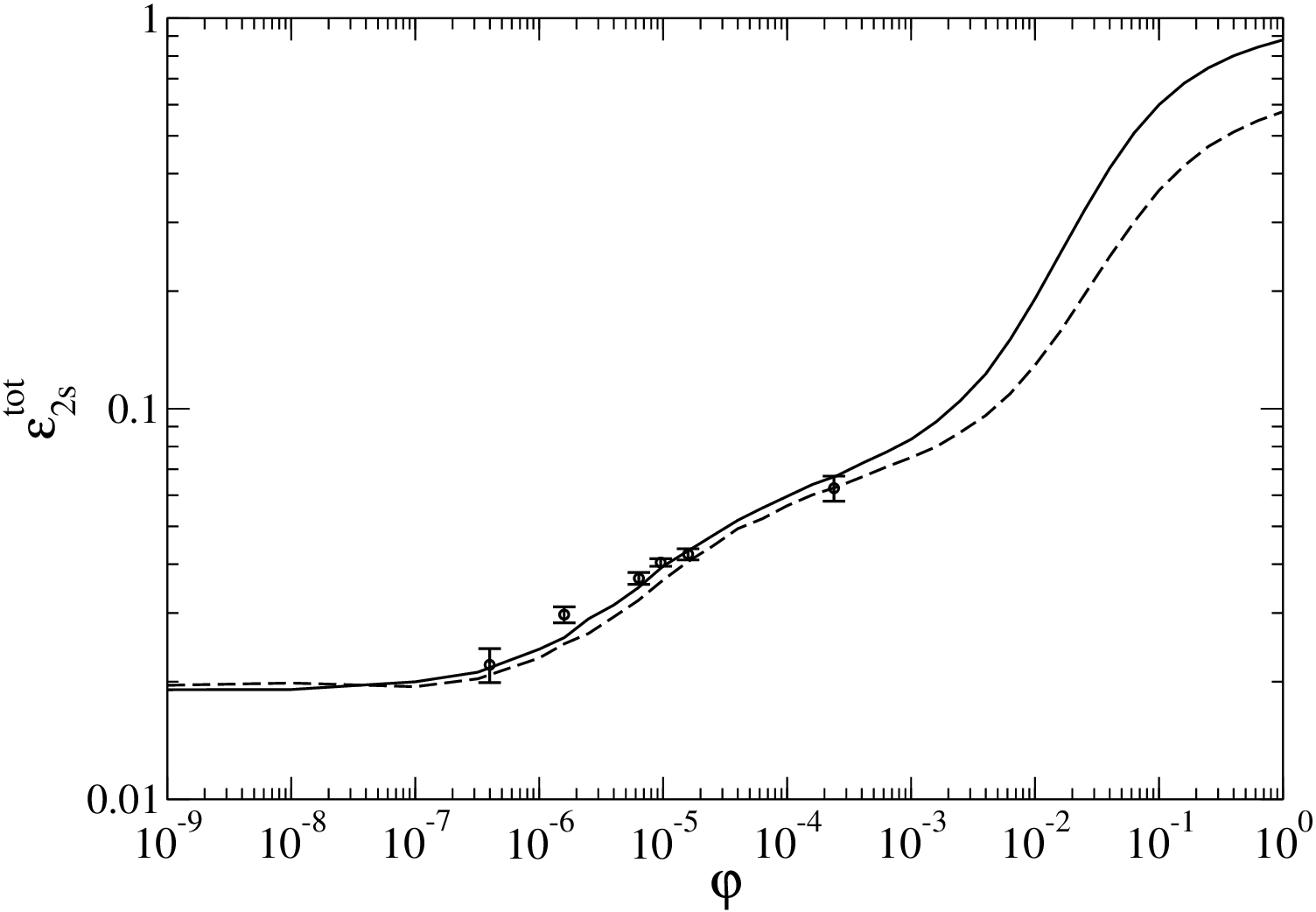}}
\caption{  
The total arrival population $\varepsilon ^{\rm tot}_{2s}$
 of $\mu^- p$ (solid line) and $\mu^- d$  
(dashed line) atoms vs the density of the target calculated 
at a target temperature $T=300$~K. The experimental data for are from \cite{HA2}.}
\label{tot_pop}
\end{figure}

The density dependence of the total arrival population for $(\mu^- p)_{2s}$ and
$(\mu^- d)_{2s}$  atoms calculated using the present version  of the atomic
cascade is shown in Fig.~\ref{tot_pop}.  Calculations were performed at relative
hydrogen  density $\varphi = (10^{-9}-1)$ and at target temperature $T=300$~K.

The total arrival populations for the $(\mu^- p)_{2s}$ and $(\mu^- d)_{2s}$ have
similar dependences on the target density. They are about 2\% in the density
range $\varphi=(10^{-9}-10^{-7})$ and grow up to 88\% and 57\% at a liquid
hydrogen density for $\mu^- p$ and $\mu^- d$ atoms, respectively.  In the case
of muonic hydrogen the density dependence of the $\varepsilon ^{\rm tot}_{2s}$
has been explained in detail in the paper \cite{PPForm}.  The same explanation
is also valid for the muonic deuterium. 

At densities below $10^{-7}$, the role of collisional processes in the
deexcitation cascade is extremely weak and the arrival population of the $2s$
state is mainly determined by $(n, l)$ distributions at the instant of the
exotic atom formation and the rates of radiative transitions from the higher
$n\,p$ states ($n\geqslant 3$).  So, in this density range, the arrival
population $\varepsilon ^{\rm tot}_{2s}$ is equal to about 2\% and is very
weakly dependent on the density for both muonic hydrogen and muonic deuterium
atoms.

In the density range $\varphi =10^{-7} - 2.5\cdot 10^{-4}$,  the arrival
population $\varepsilon^{\rm tot}_{2s}$ increases  from about 2\% up to 6.77\%
and 6.34\% for $\mu^- p$ and $\mu^- d$ atoms, respectively. In this density
range, the total population of the $2s$ state grows due to an increase of the
intensity of all collisional processes, which in turn lead to a larger
population of $np$ states ($n \geqslant 5$) and accordingly to an increase of
the intensity of the radiative $np \to 2s$ transitions.

This explanation is in accordance with the density dependence of the 
relative x-ray yields,  $Y_i=I(K_{i}/I(K_{tot})$, for $K_{i}$ lines ($i=\alpha, \beta, \gamma$, etc) (see also Figs. 17 and 18 in \cite{PPIsot}) and a strong correlation between the relative yields of $K_{i}$ lines and the total population $\varepsilon ^{\rm tot}_{2s}$.
At densities below $2.5 \cdot 10^{-4}$, the calculated arrival population $\varepsilon ^{\rm tot}_{2s}$ is in perfect agreement with the estimation based on formula (see Eq.~(21) in \cite{PPForm})
\begin{equation}
 \varepsilon^{\rm rad}_{2s} = 0.134(1- Y_{\alpha}) + 0.01 Y_{>\beta}.
 \label{Etot}
 \end{equation}
At these densities, the calculated values of $\varepsilon ^{\rm tot}_{2s}$ are in very good agreement with the experimental data obtained from the measured relative $X$-ray yields~\cite{HA2} at $H_2$ gas pressures between 0.25 and 150 Torr ($T=300$~K) that corresponds to the range of the relative densities from $3.8\cdot 10^{-7}$ up to  $2.3\cdot 10^{-4}$.

At densities above  $2.5\cdot 10^{-4}$, the populations $\varepsilon^{\rm tot}_{2s}$ and  $\varepsilon ^{\rm rad}_{2s}$ have  quite different dependencies on the density.  
The population  $\varepsilon ^{\rm rad}_{2s}$ reaches its maximum value 7.7\% and 7.5\% at $\varphi = 2\cdot 10^{-3}$  and then rapidly decreases to about 2.4\% and 1.2\% at $\varphi = 1$, respectively for $\mu^- p$ and $\mu^- d$ atoms.  Contrary to this, the arrival population $\varepsilon^{\rm tot}_{2s}$ in the density range $\varphi= 2\cdot 10^{-3} - 1$ rapidly grows from 10.4\% (for $\mu^-p$) and 8.6\% (for $\mu^-d$)   up to 88\%  and 57\% at a liquid hydrogen density  for $\mu^- p$ and $\mu^- d$ atoms, respectively. It is important to note, that these values of the total arrival  population significantly exceed the statistical weight of the $2s$ state (25\%) especially in the case of the muonic hydrogen. 

In the density range $10^{-4} \lesssim \varphi \lesssim 3\cdot 10^{-3}$, 
Stark transitions increase the population of 
$np$ states ($3\leqslant n\leqslant 5$) from which 
the radiative $np\to 2s$ transitions populate the $2s$ state.
This explanation is supported by the density dependence of the absolute 
x-ray yields of the $K\beta, K\gamma$ and $K\delta$ lines (see Fig.~15 in \cite{PPIsot}).

At a density above $\approx 3\cdot 10^{-3}$,  the CD $5\to4$, $4\to3$, and $3\to2$
together with Auger and radiative transitions result in very fast increasing of the
population $\varepsilon^{\rm tot}_{2s}$. Besides, at the density $\varphi \gtrsim
10^{-2}$  the rate of the Stark $2p \to 2s$ transition is comparable with the rate
of the radiative transition $2p\to 1s$  and becomes even about two orders of the
magnitude more at a liquid hydrogen density ($\varphi~=~1$).  Therefore, at the
density above $10^{-2}$ the initial population of the $2s$ state is mainly formed by
the  Auger transition $3p\to 2s$, the CD $3l\to2s\, (l=0, 1, 2)$, and the always
open  Stark $2p\to 2s$ transition. 

The kinetic energy of the muonic atom changes during the atomic cascade as a result
of the complex interplay of various cascade processes preceding its arrival in the
$2s$ state.    In cascade calculations, the kinetic energy distribution of exotic
atoms at the time of their transition to the $2s$ state is described by the
probability density, $\omega_{2s} (E; \varphi , T)$, which is determined by the
preceding cascade processes as well as by the target density and temperature.  
\begin{figure}[!ht]
\centerline{
\includegraphics[width=\columnwidth,keepaspectratio]
{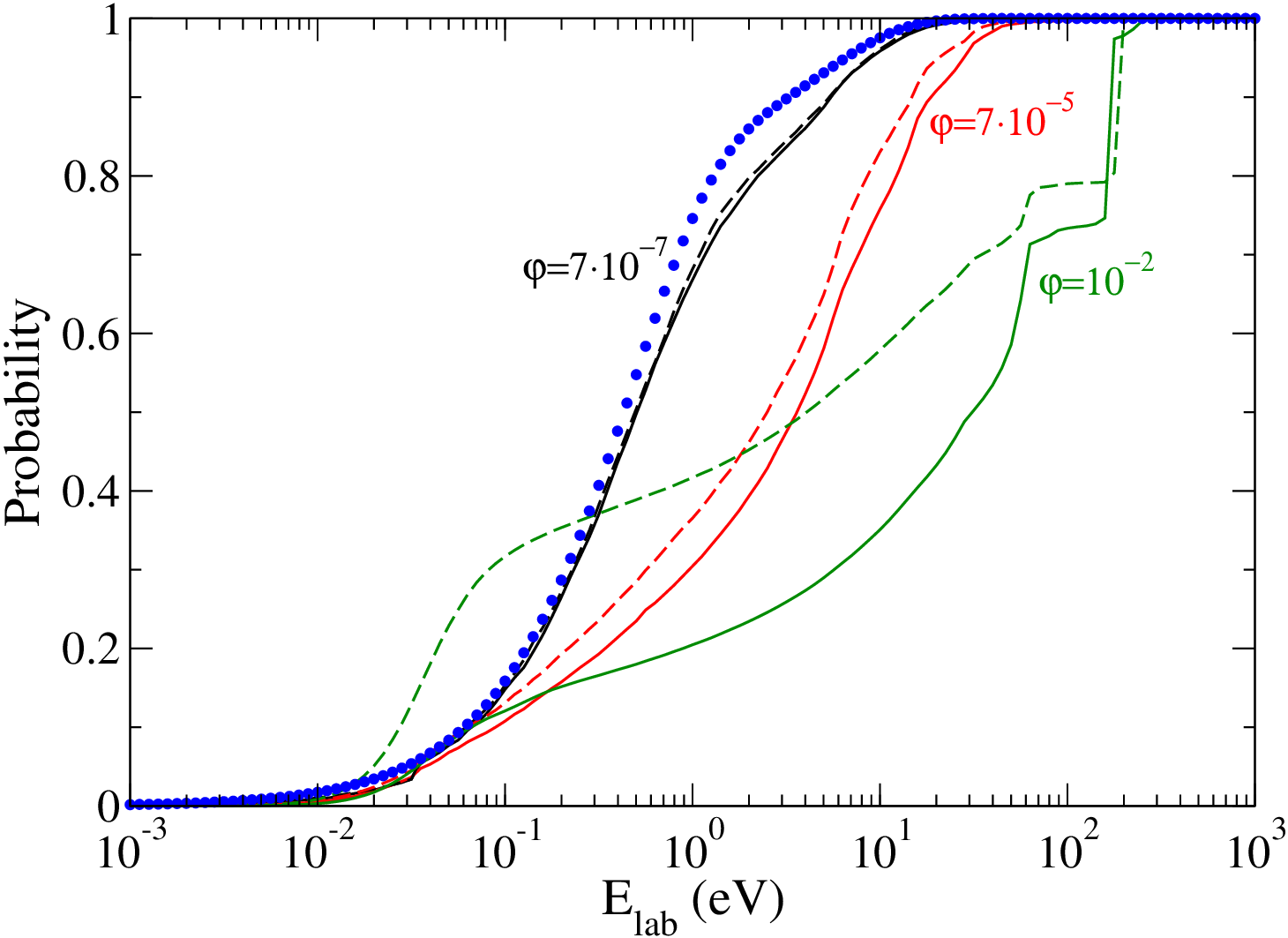}}
\caption{ Kinetic energy distributions
of muonic hydrogen (solid lines) and deuterium (dashed lines)
atoms on arrival in the $2s$ state vs. the laboratory kinetic 
energy calculated at a target temperature $T=300$~K 
for three values of the relative target density: 
$7\cdot 10^{-7}$,  $7\cdot 10^{-5}$
and $10^{-2}$. The dotted line corresponds
to the initial energy distribution of muonic atoms at the
instant of their formation (see Eq.~(\ref{initE})).}
\label{2s_dist}
\end{figure}

In Fig.~\ref{2s_dist} we present 
the distribution function of the probability,
\begin{equation}
\label{IntE}
W_{2s}(E; \varphi , T)=\int _{0}^{E}\omega_{2s}(E'; \varphi , T)dE',
\end{equation}
which determines the probability that the muonic atom has kinetic energy in the laboratory frame less than $E$ at the instant of the $2s$ state formation. 
Further, this probability will be called the kinetic energy distribution of muonic hydrogen or muonic deuterium atoms on arrival in the $2s$ state. Calculations were performed at a room target temperature $T=300$~K for three values of the relative target density: 
$\varphi = 7\cdot 10^{-7}$,  $7\cdot 10^{-5}$,  and $10^{-2}$. The dotted line in the figure shows the kinetic energy distribution of muonic atoms at the instant of their formation, given by the formula 
\begin{equation}
W_{\rm{in}}(E)=1 - \varkappa \exp(-E/E_1)-(1-\varkappa)\exp(-E/E_2),
\label{initE}
\end{equation}
where parameters $\varkappa = 0.805$, $E_1 = 0.469$~eV, and 
$E_2 = 4.822$~eV were determined in~\cite{PPForm} 
from the fit of experimental data~\cite{RPD} at the density corresponding to the target pressure $p_{\rm{H}_2} = 0.0625$ hPa and temperature $T=300$~K.

As it is seen from Fig.~\ref{2s_dist}, the kinetic energy 
distributions of the $2s$ state for both muonic hydrogen and deuterium 
atoms are practically indistinguishable at the density 
$\varphi = 7\cdot 10^{-7}$ and have a small differences from the 
initial energy distribution given by Eq.~(\ref{initE}).  At very low 
target densities $\varphi \lesssim 10^{-6}$ 
the deexcitation cascade is practically purely radiative and 
the initial kinetic energy distributions of muonic atoms at the instant 
of their formation are conserved up to the end of the cascade. 
In this density range the kinetic energy distributions 
of both muonic hydrogen and deuterium atoms in the $2s$ state 
show a very small acceleration due to the Coulomb deexcitation 
of highly-excited states.

At densities $\varphi \gtrsim10^{-6}$, the effect of the CD in the highly-excited
states (preceding the population of the $2s$ state) becomes more pronounced
providing a high-energy component in the kinetic energy distribution of the $2s$
state. In particular, at a density  $\varphi =7\cdot 10^{-5}$ (see
Fig.~\ref{2s_dist}), the contributions of the CD $8\to7$, $7\to6$, $6\to5$, and
$5\to4$ are explicitly seen in kinetic energy distributions  of both $(\mu^-
p)_{2s}$ and $(\mu^- d)_{2s} $ atoms.  The observed here small differences between
kinetic energy distributions of $(\mu^- p)_{2s}$ and $(\mu^- d)_{2s}$ atoms are
explained by the weak isotopic effect in cross sections of the CD for
highly-excited states (for details see \cite{PPIsot}).  With the further density
increasing the role of the CD becomes much more prominent.  In particular,
Figure~\ref{2s_dist} illustrates the significant isotopic effect of the CD  in the
kinetic energy distributions of the $2s$-state of muonic hydrogen and deuterium
atoms at a density $\varphi =10^{-2}$. Here the strong isotopic effect in the cross
sections and  accordingly in rates of the CD $5\to4$, $4\to 3$, and especially
$3\to 2$ (e.g., see Figs. 8, 9,11, and 12 in \cite{PPIsot}) leads to significant
differences in the kinetic energy distributions of atoms $(\mu^- p)_{2s}$ and
$(\mu^- d)_{2s}$  at the instant of their formation.

In accordance with the kinetic energy distribution of the $2s$ state 
(see Fig.~\ref{2s_dist}) the total arrival population 
$\varepsilon^{\rm tot}_{2s}$ can be divided into two fractions
\begin{equation}
\varepsilon^{\rm tot}_{2s}(\varphi) = \varepsilon^{\rm a}_{2s}(\varphi) +
\varepsilon^{\rm b}_{2s}(\varphi),
\label{pop}
\end{equation}
where $\varepsilon^{\rm a}_{2s}(\varphi)$ and 
$\varepsilon^{\rm b}_{2s}(\varphi)$
denote arrival populations of  
fractions with the kinetic energy above and below the $2s - 2p$ 
threshold, respectively. Absolute values of 
$\varepsilon^{\rm a}_{2s}(\varphi)$ 
and $\varepsilon^{\rm b}_{2s}(\varphi)$ calculated for 
$(\mu^- p)_{2s}$ and $(\mu^- d)_{2s}$ atoms at target densities
from $\varphi = 10^{-8}$ up to 1 and a target temperature $T=300$ K
are shown in Table \ref{initpop}.   
\begin{table}
\caption{The density dependence of arrival populations 
$\varepsilon^{\rm a}_{2s}$ and 
$\varepsilon^{\rm b}_{2s}$ (per formed 
muonic atom) of $(\mu^- p)_{2s}$ and $(\mu^- d)_{2s}$ fractions
above and below of the $2s - 2p$ threshold, respectively.}
\tabcolsep=0.25em
\begin{tabular} {c c c c c}
\hline \hline
,\,\,\,\,\,$\varphi$\,\,\,\,\,\,\,\,\,&\,\,\,\,\,\,\,\,\,\,\,\,\,\,\,\,\,\,\,\,\,\,\,\,\,\,$(\mu^- p)_{2s}$
&\,&\,\,\,\,\,\,\,\,\,\,\,\,\,\,\,\,\,\,\,\,\,$(\mu^- d)_{2s}$ &\\
(lhd)\,&$\varepsilon^{\rm a}_{2s}$(\%)&$\varepsilon^{\rm b}_{2s}$(\%) 
&$\varepsilon^{\rm a}_{2s}$(\%)&$\varepsilon^{\rm b}_{2s}$(\%) \\ \hline
$10^{-8}$  &   1.06  & 0.87  & 1.05   & 0.88\\
$10^{-7}$  &   1.11  & 0.90  & 1.12   & 1.02\\
$10^{-6}$  &   1.48  & 0.98  & 1.48   & 1.03\\
$10^{-5}$  &   2.82  & 1.10  & 2.57   & 1.14\\
$10^{-4}$  &   4.64  & 1.28  & 4.10   & 1.41\\
$10^{-3}$  &   6.13  & 2.21  & 4.76   & 2.69\\
$10^{-2}$  &  15.74 & 3.35  & 7.89   & 4.95\\
$10^{-1}$  &  54.67 & 5.37  & 29.05 & 7.07\\
$10^{0}$   & 76.02  &11.93  & 46.55 & 11.04\\
 \hline \hline
\end{tabular}
\label{initpop}
\end{table}

It is important to notice some peculiarities in the formation of these fractions in muonic hydrogen and muonic deuterium atoms. First, the arrival population $\varepsilon^{\rm a}_{2s}$ for the muonic hydrogen and deuterium atoms with kinetic energies above the $2s-2p$ threshold is larger than the arrival population $\varepsilon^{\rm b}_{2s}$ with kinetic energies below the $2s-2p$ threshold at all values of the target density. 
Second, the arrival population $\varepsilon^{\rm a}_{2s}$ grows with the density  much faster than the arrival population  $\varepsilon^{\rm b}_{2s}$ for both $(\mu^- p)_{2s}$ and $(\mu^- d)_{2s}$ atoms. 
Finally, the arrival population $\varepsilon^{\rm b}_{2s}$ of the muonic deuterium is greater than the arrival population of the same fraction of the muonic hydrogen at all densities of the target except for the region very close to the liquid hydrogen density. 
On the contrary,  at $\varphi >10^{-6}$, the arrival population $\varepsilon^{\rm a}_{2s}$ for the muonic hydrogen is more (and even significantly more at $\varphi \geqslant10^{-3}$) than the arrival population of the same fraction in the case of the muonic deuterium.    

 \subsection {\bf Collisional rates of $(\mu^- p)_{2s}$ 
 and $(\mu^- d)_{2s}$ atoms} 

After the formation of the hydrogen exotic atoms in the $2s$ state, their further
fate is determined by the rate of the muon decay ($\lambda_{\mu}=4.54\cdot 10^{5}$
\rm{ps}$^{-1}$) and by the rates of the collisional processes: the elastic
scattering $2s-2s$, the collision-induced radiative quenching, the Stark transition
$2s\to 2p$ followed by the fast radiative transition $2p\to 1s$, and the CD $2s\to
1s$.  The rate of the collisional process in the laboratory frame is defined as
follows
\begin{equation}
\label{rate}
\lambda_{nl \to n'l'}(E_{\rm{lab}}) = 
\varphi N_{\rm{LHD}} \sigma_{nl \to n'l'}(E_{\rm{lab}})
\sqrt{\frac{2E_{\rm{lab}}}{M}},
\end{equation}
where $E_{\rm{lab}}$  and $M$ are the kinetic energy in the 
laboratory frame and the mass of the muonic atom, respectively.        

\begin{figure}[!ht]
\centerline{\includegraphics[width=0.47\textwidth,keepaspectratio]
{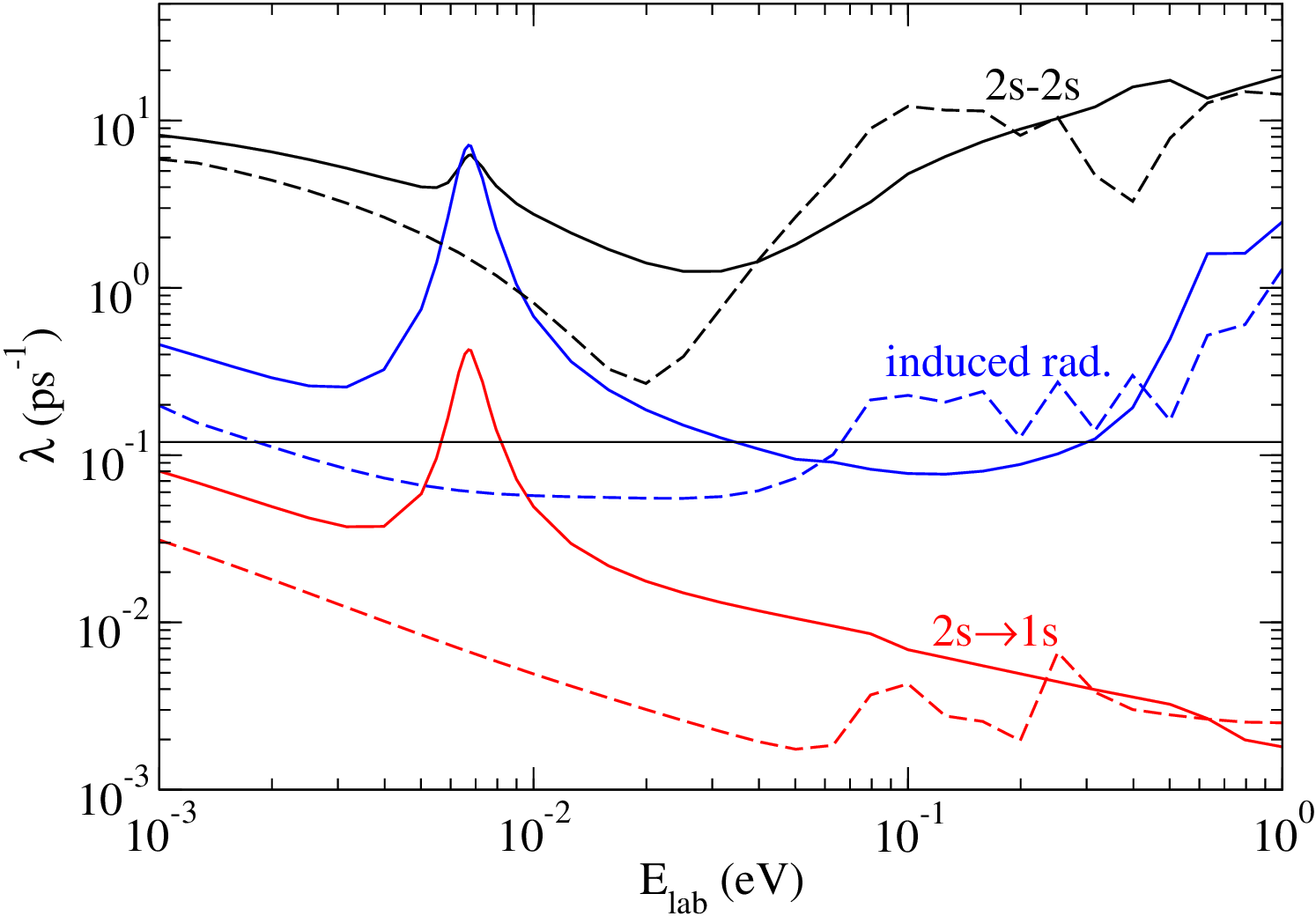}}
 \caption{ Collisional
 rates of the elastic $2s-2s$ scattering, CD 
 $2s\to1s$, and collision-induced radiative quenching of the 
 $2s$ state as a function of the laboratory kinetic energy 
 for $(\mu^- p)$ (solid lines)
 and $(\mu^- d)$ (dashed lines) atoms calculated
 at the liquid hydrogen density.}
 \label{rates_2s_nm30}
        \end{figure}

Figure~\ref{rates_2s_nm30} shows the energy dependence of collisional rates for the
elastic scattering $2s - 2s$, CD $2s \to 1s$, and collision-induced radiative
quenching of the $2s$ state for muonic hydrogen and deuterium atoms  at the liquid
hydrogen density ($\varphi=1$).   The rate of the radiative transition $\lambda
^{rad}(2p\to 1s)=0.12$ \rm{ps}$^{-1}$ is shown for comparison.   A small difference
of about 5\% between radiative rates of the $2p\to 1s$ transition in muonic
hydrogen and deuterium due to the difference of their reduced masses is not visible
at the scale of the figure. 
        
The isotope effect observed in cross sections for scattering of muonic atoms in
states with values of the principal quantum number $n < 8$ (see \cite{PPIsot}) is
practically absent for highly excited states with $n > 8$.  The only source of the
isotopic effect in the kinetics of the atomic cascade for states with $n>8$ is the
trivial kinematic factor $\sqrt{2E_{\rm{lab}}/M}$ in Eq.~(\ref{rate}). In the case
of  the $2s$ state, the isotopic effect revealed in cross sections (see
Figs.~\ref{crsec_mup}  and \ref{crsec_mud_2s}) is additionally enhanced by this
factor associated with the difference of the exotic atom velocities at the same
energy in the laboratory system.  It is seen in Fig.~\ref{rates_2s_nm30}, that the
isotopic  effect in rates of all collisional processes is very  strong.  

 \subsection {\bf Intensities of the different disintegration modes
 of $(\mu^- p)_{2s}$ and $(\mu^- d)_{2s}$ } 

\begin{figure}[!ht]          
\centerline{
\includegraphics[width=\columnwidth,keepaspectratio]
{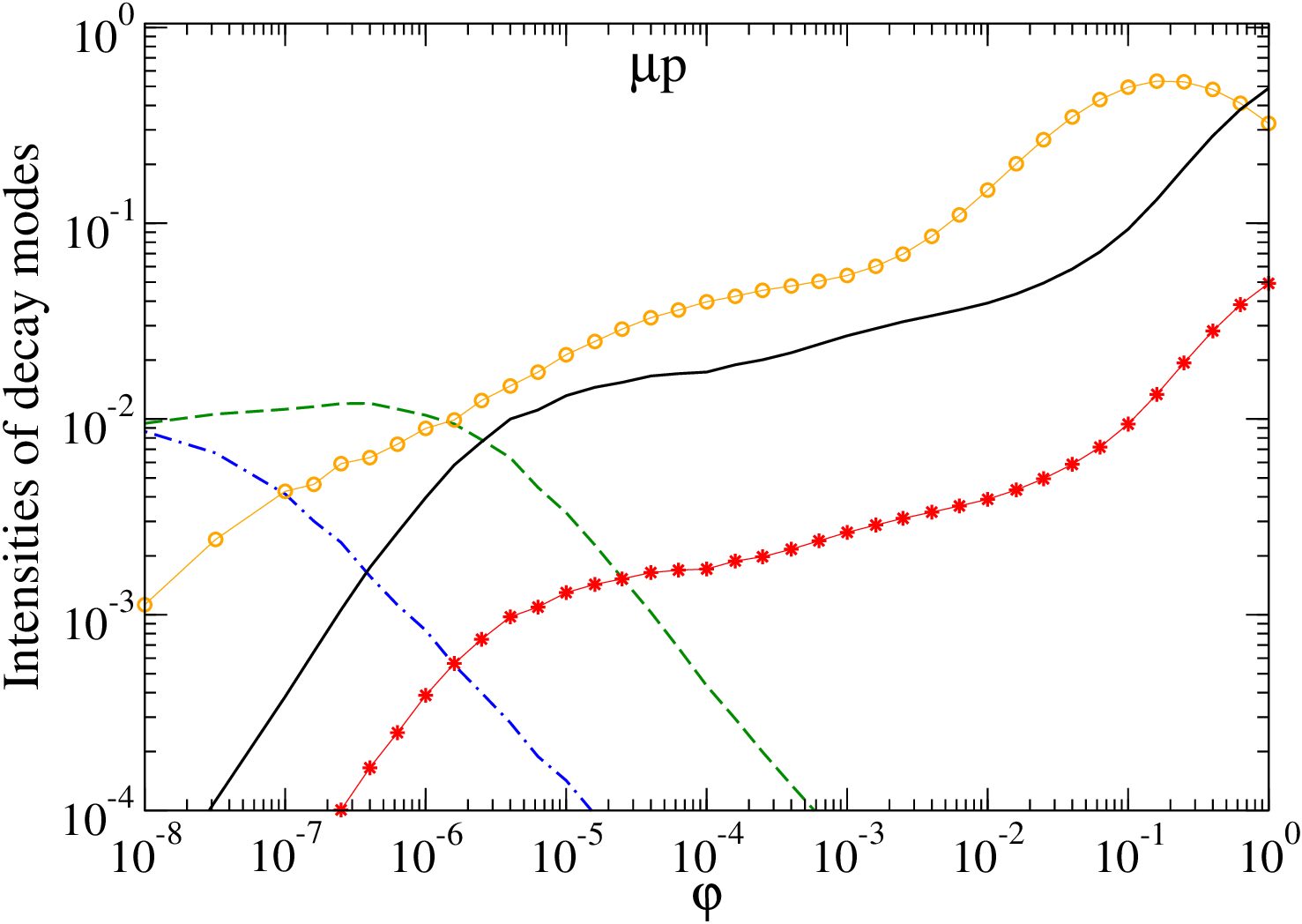}}
\caption{Intensities (per one formed $\mu^- p$ atom) 
for all disintegration modes  
of $(\mu^- p)_{2s}$ vs the relative target density:
collision-induced radiative quenching (solid line), 
Stark $2s\to 2p$ transition (line with open circles), 
Coulomb de-excitation (line with asterisks), and
$\mu$-decay (dashed and dashed-dotted lines  
for fractions below and above the $2s-2p$ threshold, respectively).}
\label{quenmup2s300K}
\end{figure}

\begin{figure}[!ht]          
\centerline{
\includegraphics[width=\columnwidth,keepaspectratio]
{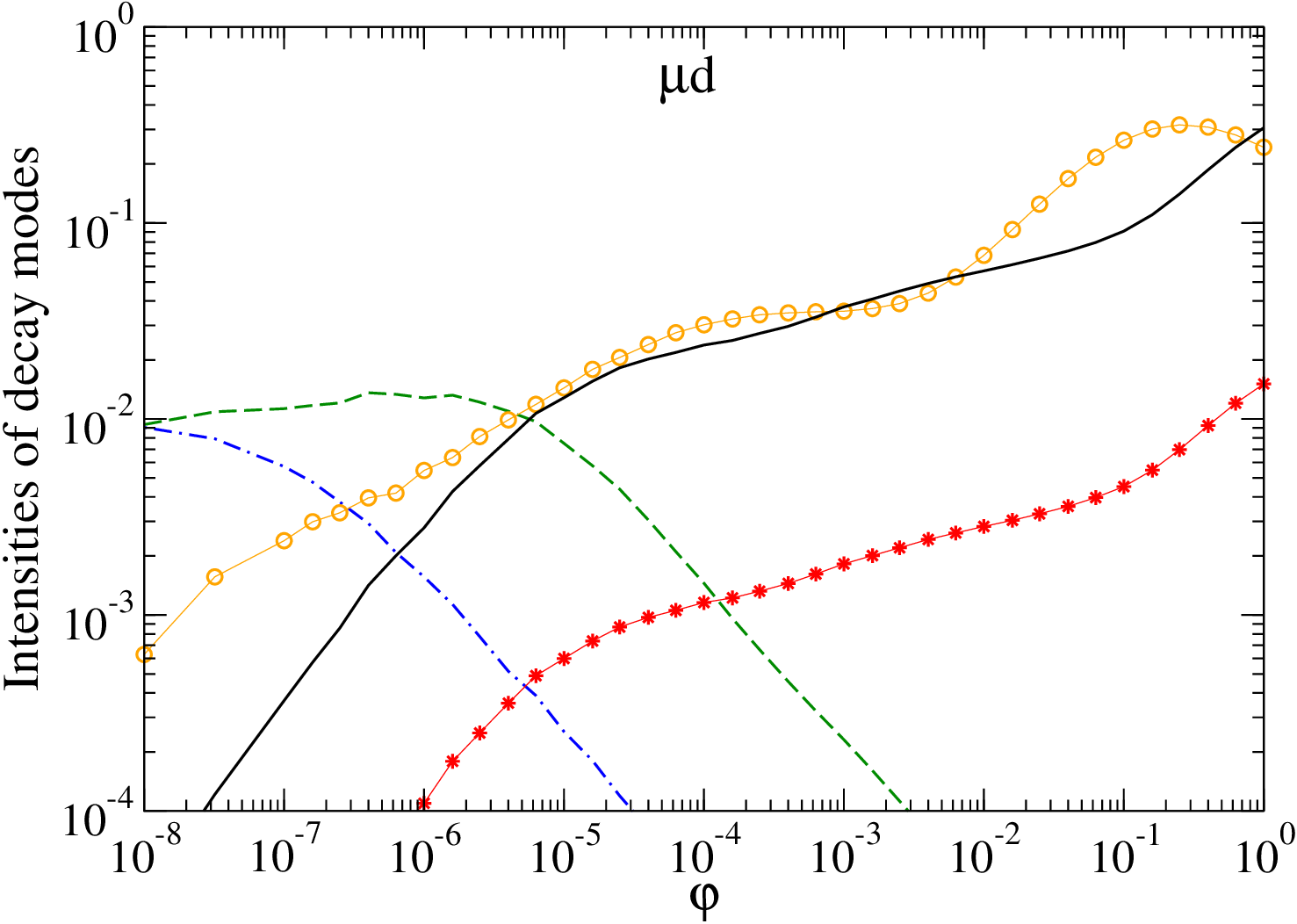}}
\caption{  The same as in Fig.~\ref {quenmup2s300K} for 
$(\mu^- d)_{2s} + D$ collisions.}
\label{quenmud2s300K}
\end{figure}

The intensities of the decay modes, i.e. the number of atoms (per one formed muonic atom) disintegrated through a given mode were calculated within the present cascade model for muonic hydrogen and deuterium atoms in $2s$ state. 
The intensities of the destruction of the $2s$ state for  $(\mu^- p)$ and $(\mu^- d)$ atoms are shown as functions of the target density in Figs. \ref{quenmup2s300K} and \ref{quenmud2s300K}, respectively, and are also presented in Tables \ref{decmup} and \ref{mupdcomp} for several  density values. 

The figures and tables show the intensities of the following destruction modes:\\
$I_{\rm rqu}$ -- collision-induced radiative quenching,\\
$I_{\rm Str}$ -- Stark transition $2s\to 2p$ followed by the fast
radiative $2p\to 1s$ transition,\\
$I_{\rm CD}$ -- Coulomb deexcitation $2s\to 1s$,\\
$I^{\rm a}_{\mu}$ --  $\mu^-$-decay at kinetic energies 
above the $2s-2p$ threshold,\\
$I^{\rm b}_{\mu}$ --  $\mu^-$-decay at kinetic energies below the $2s-2p$ threshold.\\

At the lowest densities,  the muon decay is the main process resulting in the
disintegration of the $2s$ state of both muonic hydrogen and deuterium atoms at kinetic
energies above and below the $2s-2p$ threshold (see also Table II). 

At kinetic energy above the $2s-2p$ threshold, the intensity of the $\mu$-decay
decreases rather rapidly with the density increasing. On the contrary, the intensity of
$\mu$-decay at kinetic energies below the $2s-2p$ threshold at first even increases
with increasing density (especially in the case of  $(\mu^- d )_{2s}$) and  then
decreases rapidly with further increase in density at $\varphi > 10^{-5}$.

\begin{table}
\caption{The density dependence of intensities (per one formed 
muonic atom, in percent) of the $\mu$-decay calculated for $(\mu^- p)_{2s}$
and $(\mu^- d )_{2s}$ with kinetic energies both above and below 
the $2s - 2p$ threshold ($I^{\rm a}_{\rm w}$ and $I^{\rm b}_{\rm w}$, 
respectively).}
\tabcolsep=0.25em
\begin{tabular} {c c c c c}
\hline \hline
\,\,\,\,\,$\varphi$\,\,\,\,\,\,\,\,\,\,\,&\,\,\,\,\,\,\,\,\,\,\,\,\,\,\,\,\,\,\,\,\,\,\,\,\,\,$(\mu^- p)_{2s}$
&\,\,\,\,\,\,\,\,\,\,\,\,\,&\,\,\,\,\,\,\,\,\,\,\,\,\,\,\,\,\,\,\,\,\,\,\,\,\,$(\mu^- d)_{2s}$&\,\,\\
(lhd)\,&\,\,\,\,\,$I^{\rm a}_{\rm w}$\,\,\,\,&\,\,\,$I^{\rm b}_{\rm w}$\,\,\,&
\,\,\,\,\,\,$I^{\rm a}_{\rm w}$\,\,\,\, &\,\,\,\,\,$I^{\rm b}_{\rm w}$\,\,\,\,\,\, \\ \hline
$10^{-8}$ &0.86   &0.95   & 0.91    & 0.95 \\
$10^{-7}$ &0.41   &1.12   & 0.57    & 1.02 \\
$10^{-6}$ &0.08   &1.05   & 0.16    & 1.43 \\
$10^{-5}$ & 0.02  &0.33   & 0.03    & 0.79 \\
$10^{-4}$ & 0.00  &0.04   & 0.00    & 0.14 \\
$10^{-3}$ & 0.00  &0.01   & 0.00    & 0.02 \\
 \hline \hline
\end{tabular}
\label{decmup}
\end{table}

\begin{table}
\caption{The density dependence of intensities
$I_{\rm CD}, I_{\rm rqu},$ and $I_{\rm Str}$ ( in percent) calculated
for $(\mu^- p)_{2s}$ and $(\mu^- d)_{2s}$.}
\tabcolsep=0.25em
\begin{tabular} {c c c c c c c}
\hline \hline
\,$\varphi$\,\,\,\,\,\,&\,\,\,\,\,\,\,\,\,&\,\,$(\mu^- p)_{2s}$
\,&\,\,\,\,\,\,\,\,\,\,\,\,\,\,\,\,\,\,\,\,&\,\,\,\,\,\,\,&$(\mu^- d)_{2s}$&\,\,\,\,\,\,\,\\
(lhd)\,\,\,\,\,\,\,\,\,&\,\,\,\,$I_{\rm CD}\,\,$&\,
$I_{\rm rqu}$\,\,&\,$I_{\rm Str}$\,\,&\,\,\,\,\,$I_{\rm CD}$\,\,&\,\,\,\,\,
$I_{\rm rqu}$\,\,\,&\,\,\,\,\,$I_{\rm Str}$\,\, \\ \hline
$10^{-8}$  &   0.00    & 0.00 & 0.11&0.00  & 0.004 & 0.07\\
$10^{-7}$  &   0.004    & 0.04 & 0.43&0.00  & 0.04 & 0.27\\
$10^{-6}$  &   0.04    & 0.40 & 0.90&0.01  & 0.31 & 0.58\\
$10^{-5}$  &   0.13    & 1.32 & 2.13&0.06  & 1.36 & 1.48\\
$10^{-4}$  &   0.17    & 1.74 & 3.97&0.11  & 2.25 & 3.01\\
$10^{-3}$  &   0.26    & 2.66 & 5.41&0.18  & 3.70 & 3.55\\
$10^{-2}$  &  0.39    & 3.91 & 14.76&0.28  & 5.69 & 6.85\\
$10^{-1}$  &  0.94    & 9.33 & 49.52&0.45  & 9.09 & 26.45\\
$10^{0}$   & 4.93    & 48.98 & 32.38&1.51  & 30.59 & 24.33\\
 \hline \hline
\end{tabular}
\label{mupdcomp}
\end{table}

The intensities of Stark $2s\to 2p$ transitions in muonic hydrogen and deuterium atoms are in accordance with the primary populations of these atoms at kinetic energies above the $2s-2p$ threshold but turn out to be less than the values of the corresponding primary populations. 
The intensity of the Coulomb deexcitation for $(\mu^- p)_{2s}$ are about 1.5 - 2 times higher than in the case of $(\mu^- d)_{2s}$, while the values of the radiation quenching intensity for these two isotopes are very close at the target density less than $10^{-5}$ and with increasing density targets up to $10^{-2}$ this ratio changes in favor of the muonic deuterium atom by about 1.5 times. 

The fastest collisional process -- the elastic $2s-2s$ scattering -- leads to a deceleration of muonic atoms in the $2s$ state. 
As a result,  some part of the muonic $2s$-state atoms initially formed with kinetic energies above the $2s-2p$ threshold turns out to be below the threshold.
Thus, the population of the fraction with kinetic energy above the $2s-2p$ threshold decreases, and the population of the fraction with kinetic energy below the $2s-2p$ threshold increases accordingly compared to their values at the instant of formation.
Thus, the total population of the fraction with kinetic energy below the $2s-2p$ threshold is always greater than its initial population $\varepsilon^{\rm b}_{2s}$. 
 This influence of elastic scattering on redistribution of fractions with kinetic energies above and below the $3s-2p$ threshold (in favor of increasing the sub-threshold fraction) increases with increasing target density.

In Table \ref {quless} we present the initial population $\varepsilon^{\rm b}_{2s}$ and the total number of the disintegrated $(\mu^- p)_{2s}$ atoms with the energy below the $2s-2p$ threshold $I^{\rm b}_{2s}$ (per one formed $\mu^- p$ atom):
\begin{equation}
I^{\rm b}_{2s} = I^{\rm b}_{\mu}+I_{\rm CD}+I_{\rm rqu}
\label{qubel}
\end{equation}
in comparison with the experimental data \cite{RPD, PDH}. 
\begin{table}[h!]
\caption{The initial population $\varepsilon^{\rm b}_{2s}$ and the 
total number of the disintegrated $(\mu^- p)_{2s}$ atoms with the energy below the $2s-2p$ threshold $I^{\rm b}_{2s}$ (per one formed 
$\mu p$ atom).}
\tabcolsep=0.25em
\begin{tabular} {c c c c c}
\hline \hline
$\varphi$&\,\,\,$\varepsilon^{\rm b}_{2s}$(\%) &\,\,\,&
$I^{\rm b}_{2s}$ (\%)\,\,\,\,\,\,\,\,\,\,\,\,\,\,\,\,\,\,\,\,\,\,\,\,\,\,\,&\\
(lhd)  &\,\,\,Theory\,\,\,\,&\,\,\,\,\,\,\,Theory &\,\,\,\,\,\,\,\,\,\,Exp. \,\,\,\,\,\,\,& \,\,Ref. \\ \hline
$7.35\cdot 10^{-8}$ &0.90 &    1.12& $ 0.85 ^{+0.14}_{-0.14} $& \cite{RPD}\\
$2.94 \cdot 10^{-7}$ &0.95 &   1.27& $ 0.92 ^{+ 0.08}_{-0.08} $& \cite{RPD}\\
$1.18 \cdot 10^{-6}$ &1.05 &    1.51& $ 1.10 ^{+ 0.08}_{-0.08} $& \cite{RPD}\\
$4.71 \cdot 10^{-6}$ & 1.13 &   1.72& $ 1.25 ^{+ 0.98}_{-0.48} $& \cite{PDH}\\
$1.88\cdot 10^{-5}$ &  1.15 &   1.84& $ 1.07 ^{+ 0.57}_{-0.30} $& \cite{PDH}\\
$7.53\cdot 10^{-5}$ &  1.28 &   1.93& $ 1.63 ^{+ 1.08}_{-0.35} $& \cite{PDH}\\
 \hline \hline
\end{tabular}
\label{quless}
\end{table}   

On the whole, there is a satisfactory agreement between the theoretical results and
experimental data.  An exception is the experimental value obtained at a target density
of $1.88\cdot 10^{-6}$ \cite{PDH}, which is much less the corresponding theoretical
value. It should be noted that this experimental value falls out of the general trend
of population growth with increasing target density and, moreover, there are no
arguments to explain it.

\subsection{Absolute x-ray yield of $K\alpha$ line }

The absolute x-ray yield of the $K\alpha$ line is determined
as the intensity of this line per muonic atom formed.
The total intensity, $I_{K\alpha}$, of the $K\alpha$ line can be 
represented as a sum of three different contributions 
corresponding to the preceding history of the atoms involved in its
formation:
\begin{equation}
I_{K\alpha}=I_{2p\to1s} + I_{\rm Str}+I_{\rm rqu}.
\label{absyield}
\end{equation}
Here $I_{2p\to1s}$ denotes the intensity of the direct $2p\to1s$ radiative transition occurring after the transition of a muonic atom to the $2p$ state from states with the principal quantum number $n\geqslant 3$. 
Remind, that  $I_{\rm Str}$ and $I_{\rm rqu}$ are the intensities of the $2s\to 2p$ Stark transition (at kinetic energies above the $2s-2p$ threshold) followed by the $2p\to1s$ radiative transition and the collision-induced radiative quenching of the $2s$ state (at kinetic energies below the $2s-2p$ threshold), respectively.
 
\begin{figure}[!ht]          
\centerline{
\includegraphics[width=\columnwidth,keepaspectratio]
{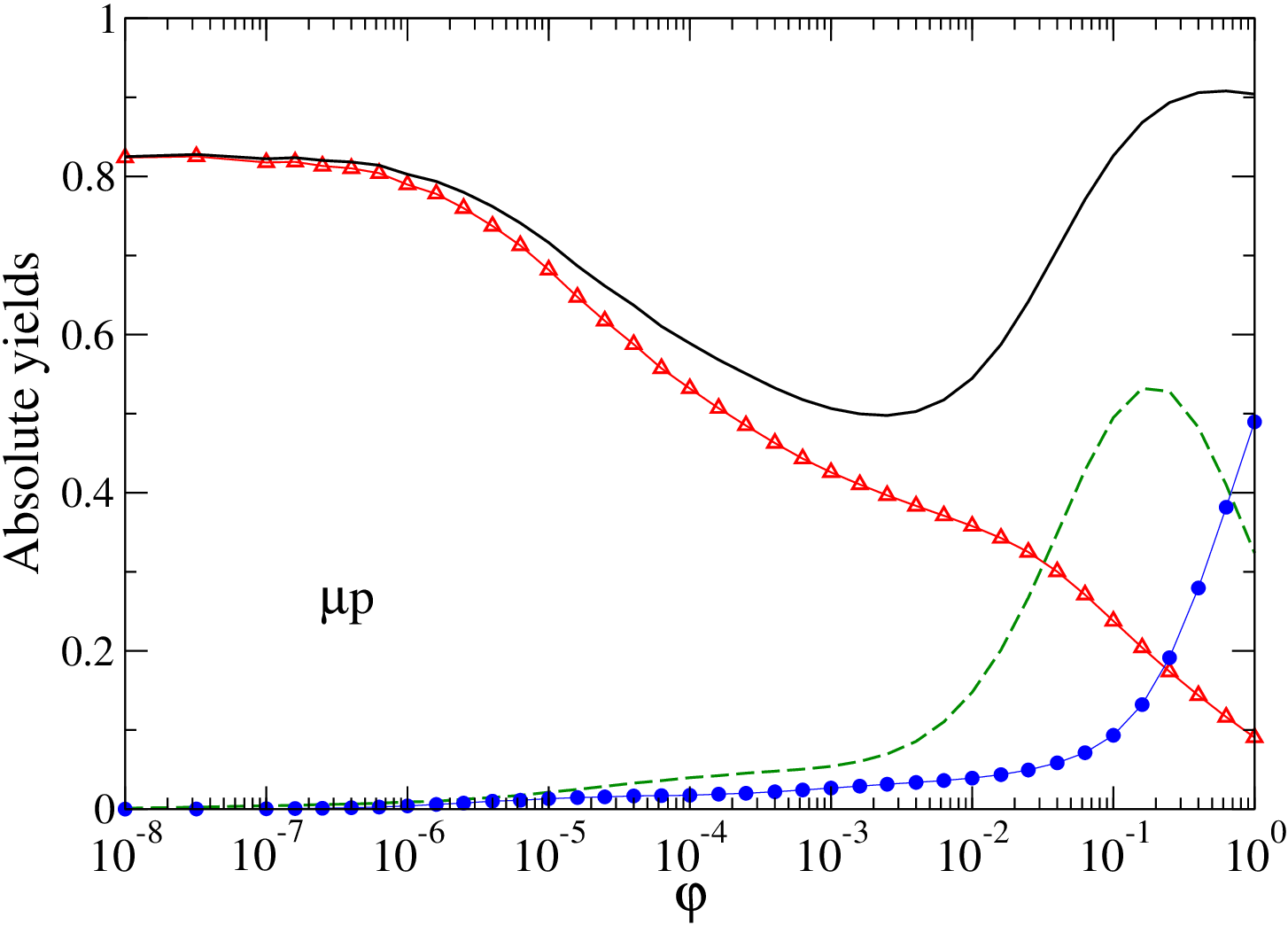}}
\caption{  The density dependence of the different
contributions to the absolute
$K\alpha$-line yield calculated for muonic hydrogen at the target temperature
$T=300$~K.: 
$I_{2p\to1s}$ (solid line with triangles), $I_{\rm Str}$ 
(dashed line), $I_{\rm rqu}$ (solid line with filled circles), 
and their sum (solid line).}
\label{absyieldmup2}
\end{figure}

\begin{figure}[!ht]          
\centerline{
\includegraphics[width=\columnwidth,keepaspectratio]
{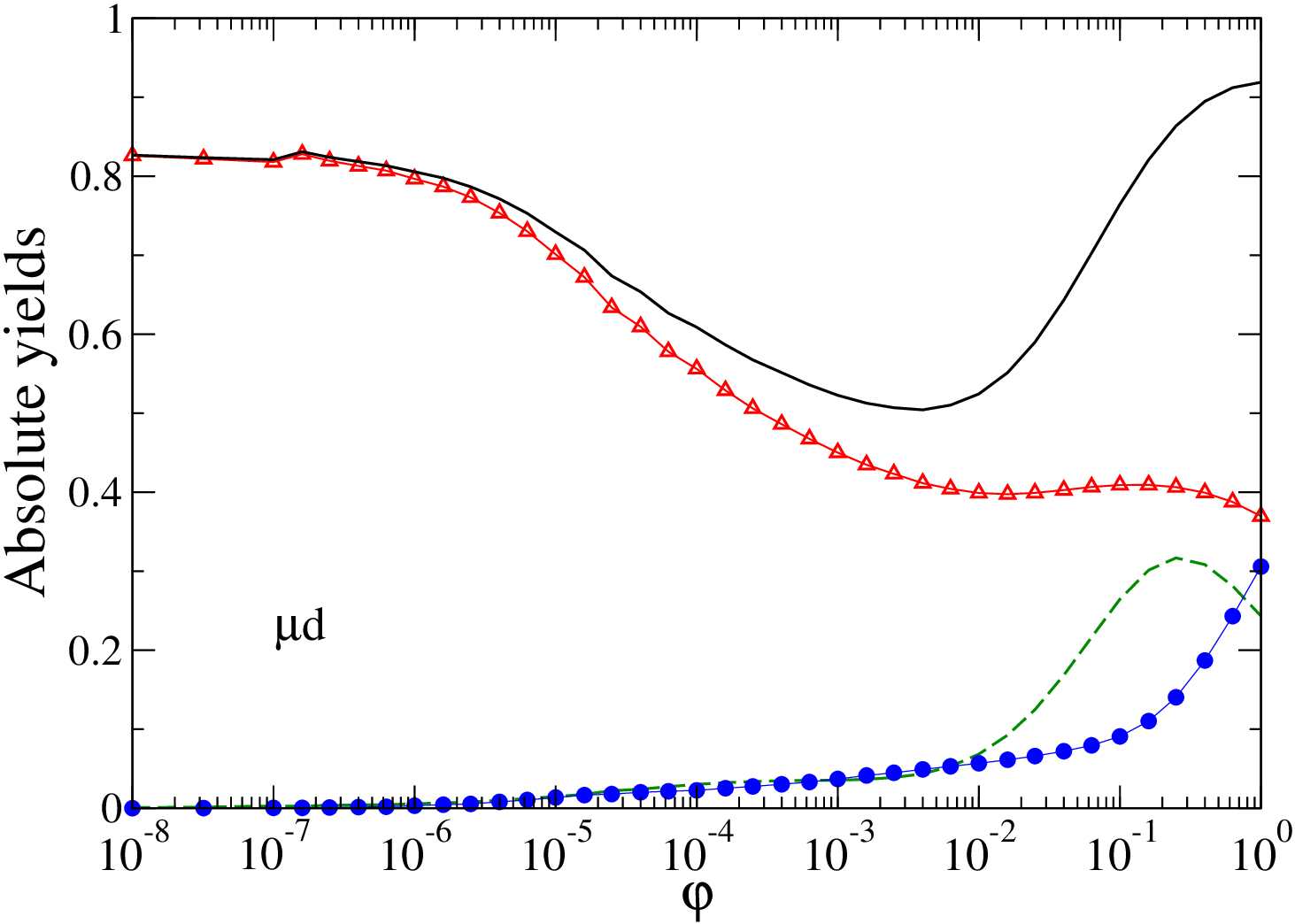}}
\caption{  The same as in Fig.~\ref {absyieldmup2}
for $\mu^- d$ atom.}
\label{absyieldmud2}
\end{figure}

Figures~\ref{absyieldmup2} and \ref{absyieldmud2} show the 
density dependence of the different modes forming the total absolute yield of the $K\alpha$ line as well as their sum calculated at a target temperature $T=300$~K for $\mu^- p$ and $\mu^- d$ atoms, respectively. 

As can be seen from Figures~\ref{absyieldmup2} and \ref{absyieldmud2}, the contribution
of the direct radiation $2p-1s$ transition $I_{2p\to1s}$ to the absolute yield of the $K\alpha$-line
decreases with increasing target density $\varphi > 10^{-6}$. In the case of a muonic
hydrogen atom, this decrease becomes rather rapid at $\varphi > 10^{-3}$ and the
contribution reaches a value of 0.1 at the density of liquid hydrogen.  In the case of
a muonic deuterium atom, we observe a similar dependence on the density at $\varphi <
10^{-3}$, however a further increase in the target density does not lead to a sharp
decrease of the contribution  $I_{2p\to1s}$ (at the density of liquid hydrogen, this contribution is about 0.4).  This difference in the dependence of $I_{2p\to1s}$  on the density in muonic hydrogen and  muonic deuterium
is explained by isotope effect in the rates of collisional processes of these atoms
(see \cite {PPIsot}) in the upper part of the cascade of deexcitation ($n\geqslant 3$).
In particular, this is reflected in the primary populations of the $2s$ state at
kinetic energies above the $2s-2p$ threshold (see Table \ref {initpop}).

\subsection {\bf Probabilities of different disintegration modes and 
lifetime of $(\mu^- p)_{2s}$ and $(\mu^- d)_{2s}$}

The  probability that the muonic atom formed in the $2s$ state disintegrates 
due to one of decay modes, $P_{i}(\varphi)$, is given by  
\begin{equation}
\label{prob}
P_{i}(\varphi)=\frac{1}{p_0}\Big\{1-\exp\left[-\int_{0}^{\infty}
\frac{\lambda_{i}(E; \varphi)}
{\lambda_{\rm tot}(E; \varphi)}F_{i}(E; \varphi)dE\right]\Big\}. 
\end{equation}
Here, the $\lambda_{i}(E; \varphi)$ is the rate of the  process $i$ where index $i$ denotes the following disintegration channels:
$i=\mu$ ($\mu$-decay), $i=\rm CD$ (Coulomb deexcitation $2s\to 1s$),
$i=\rm rqu$ (collision-induced radiative quenching), and $i= \rm Str$
(Stark transition $2s\to 2p$ resulting in the radiative transition 
$2p\to1s$). $\lambda_{\rm tot}(E; \varphi)$ is a summary rate of all
processes, which lead to the disintegration of the $2s$ state: 
\begin{equation}
\lambda_{\rm tot}(E; \varphi)=\lambda_{\mu}+ \lambda_{\rm CD}
+\lambda_{\rm rqu}+\lambda_{\rm Str}.
\label{colrate}
\end{equation}
The factor $p_0=1-e^{-1}$ provides the correct behavior of the probability  $P_{\mu}$ in the limit $\varphi \to 0$.
The kinetic energy distribution $F_{i}(E; \varphi)$ is defined at the instant of the disintegration of the $2s$ state through  the i-th decay mode and normalized by the condition:
\begin{equation} \label{Norm}
\int_0^{\infty}F_{i}(E; \varphi)dE =1.	
\end{equation}

\begin{figure}[!ht]          
\centerline{
\includegraphics[width=\columnwidth,keepaspectratio]
{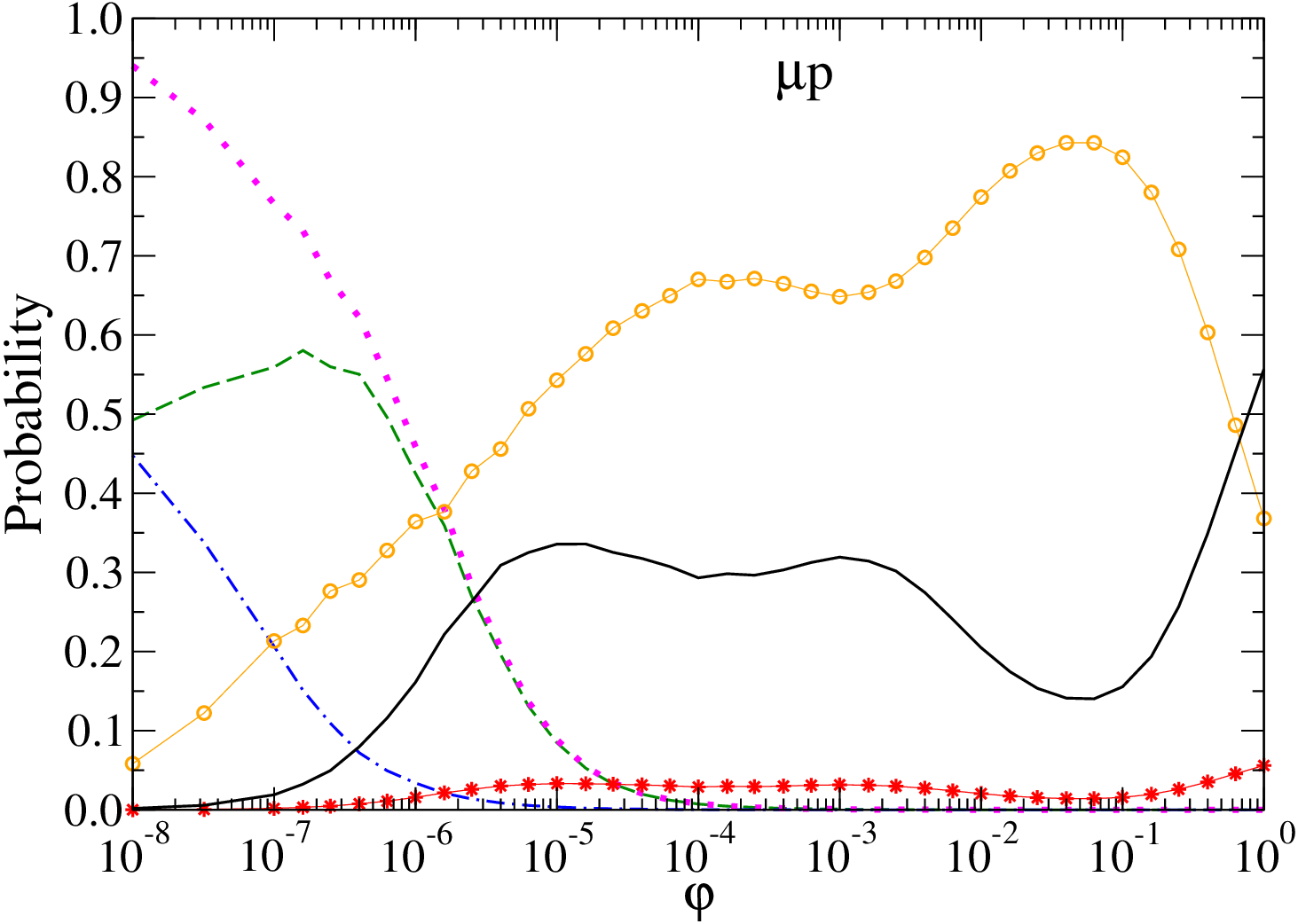}}
\caption{  The density dependence of decay probabilities 
of $(\mu^- p)_{2s}$ for various disintegration modes:
collision-induced radiative quenching (solid line), 
Stark $2s\to 2p$ transition (line with open circles), 
Coulomb de-excitation (line with asterisks), and
$\mu$-decay (dashed and dashed-dotted, and dotted lines  
for fractions below and above the $2s-2p$ threshold and their sum, 
respectively).}
\label{probmup2s300K}
\end{figure}

\begin{figure}[!ht]          
\centerline{
\includegraphics[width=\columnwidth,keepaspectratio]
{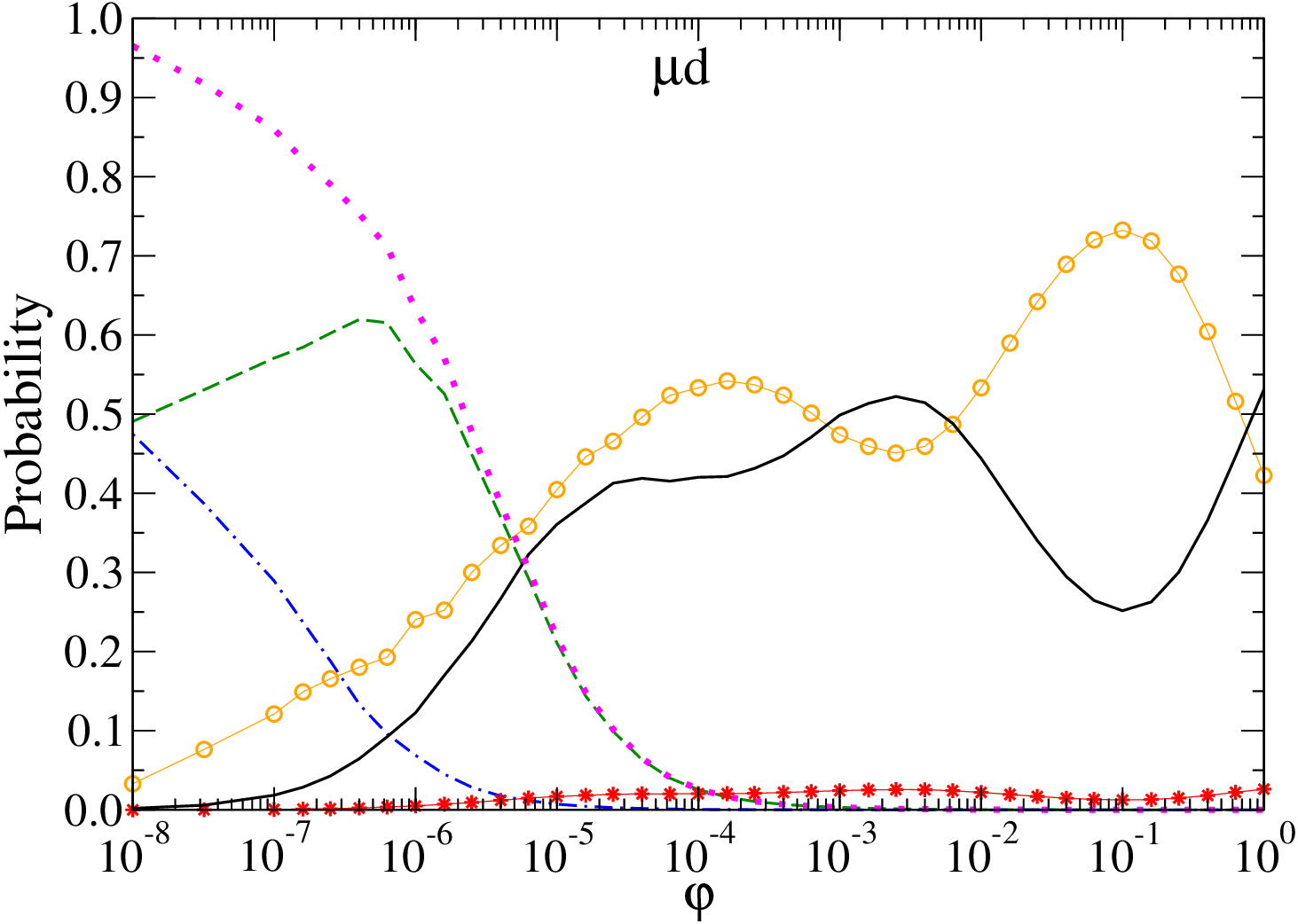}}
\caption{  The same as in Fig.~\ref {probmup2s300K} for 
$(\mu^- d)_{2s}$ atom.}
\label{probmud2s300K}
\end{figure}

The probability of the different disintegration modes for  $(\mu^- p)_{2s}$ and $(\mu^- d)_{2s}$ atoms as functions on target density is shown in Figures \ref{probmup2s300K} and \ref{probmud2s300K}, correspondingly.
As can be seen from the figures, probabilities of $\mu$-decay have a similar density dependence for muonic hydrogen and deuterium atoms. This is explained by the fact that the $\mu$-decay rate does not depend on either the energy or the target density, and the kinetic energy distributions of muonic hydrogen and deuterium are very close in the density region where the probability of muon decay is not negligible (see Fig. \ref{2s_dist}). 

On the contrary, for the probabilities of the destruction of the $2s$ state in collisional processes, we observe significant differences between muonic hydrogen and deuterium in both the values of these probabilities and their dependence on the target density.
These differences are especially pronounced in the ratio of the probabilities of Stark transitions and collision-induced radiation quenching. The observed differences are explained by two factors: significant differences in the ratio of velocities, as well as in the distributions over the kinetic energies of muonic hydrogen and deuterium atoms.

The mean time interval between the formation of the muonic atom in 
the $2s$ state and its decay we define as the
lifetime of the $2s$ state. In this work, to determine the lifetime of the $2s$ state, we use the $\mu$- decay probability
\begin{equation}
\label{mudecay}
P_{\mu }(\varphi)=\frac{1}{p_0}
\left[1-\exp \left(-\lambda_{\mu }\tau_{2s}(\varphi)\right)\right], 
\end{equation}
where a lifetime $\tau_{2s}(\varphi)$ of the $2s$ state is defined by
\begin{equation}
\label{life2s}
\tau_{2s}(\varphi)=\int_{0}^{\infty}\frac{F_{\mu }(E; \varphi)}
{\lambda_{\rm tot}(E; \varphi)}dE
\end{equation}
From Eqs.~(\ref{mudecay}) and (\ref{life2s}) we obtain 
\begin{equation}
\label{time2s}
\tau_{2s}(\varphi)=-\tau_{\mu }\ln\left[1-p_0\cdot P_{\mu}(\varphi)\right].
\end{equation}

Formulas similar to Eqs.~(\ref{mudecay})-(\ref{time2s}) allow to define and to calculate the lifetime $\tau_{2s}^{\rm a}$ for the fraction of atoms in the $2s$ state with the kinetic energy above the $2s-2p$ threshold and also the lifetime $\tau_{2s}^{\rm b}$ for the fraction of atoms in the $2s$ state with the kinetic energy below the $2s-2p$ threshold.

\begin{equation}
\label{mudec}
P_{\mu}^{\alpha}(\varphi)=\frac{1}{p_0}
\left[1-\exp \left(-\lambda_{\mu}\tau_{2s}^{\alpha}(\varphi)\right)\right], 
\end{equation}
where 
\begin{equation}
\label{life2salph}
\tau_{2s}^{\alpha}(\varphi)=\int_{0}^{\infty}\frac{F_{\mu}^{\alpha}
(E; \varphi)}{\lambda^{\alpha}(E; \varphi)}dE.
\end{equation}
is the lifetime of the fraction of the $2s$ state with the kinetic energy 
above ($\alpha = \rm a$) 
or below ($\alpha =\rm b$) the $2s-2p$ threshold, respectively.  
The corresponding total rates of these fractions are given by:
\begin{equation}
\lambda^{\rm a}(E; \varphi)=\lambda_{\mu}+\lambda_{\rm CD}(E; \varphi)
+\lambda_{\rm Str}(E; \varphi)
\label{rate_a}
\end{equation}
and
\begin{equation}
\lambda^{\rm b}(E; \varphi)=\lambda_{\mu}+ \lambda_{\rm CD}(E; \varphi)
+\lambda_{\rm rqu}(E; \varphi).
\label{rate_b}
\end{equation} 
Infinite limits of integration in Eq.~(\ref{life2salph}) are of a formal nature, 
since the probability densities $F_{\mu}^{\rm a}(E; \varphi)$ and $F_{\mu}^{\rm b}(E; \varphi) $are different from zero only in the intervals 
$E \geqslant E_{\rm thr}$ and $E\leqslant E_{\rm thr}$, respectively.
From Eqs.~(\ref{mudec}) and (\ref{life2salph}) we obtain 
\begin{equation}
\label{time2salph}
\tau_{2s}^{\alpha}(\varphi)=-\tau_{\mu}\ln\left[1-p_0\cdot P_{\mu}^{ \alpha}
(\varphi)\right].
\end{equation}

In the present study  
three variants of the $(\mu^- p)_{2s}$ and $(\mu^- d)_{2s}$ lifetime were
calculated: the lifetime, $\tau_{2s}^{\rm a}$, of the fraction with the kinetic energy 
above the $2s-2p$ threshold, the lifetime, $\tau_{2s}^{\rm b}$, of the fraction 
with the kinetic energy below the $2s-2p$ threshold, and the 
lifetime, $\tau_{2s}$, of the $2s$ state independently whether the 
kinetic energy of muonic atom is above or below the $2s-2p$ threshold. 

\begin{figure}[!ht]          
\centerline{
\includegraphics[width=\columnwidth,keepaspectratio]
{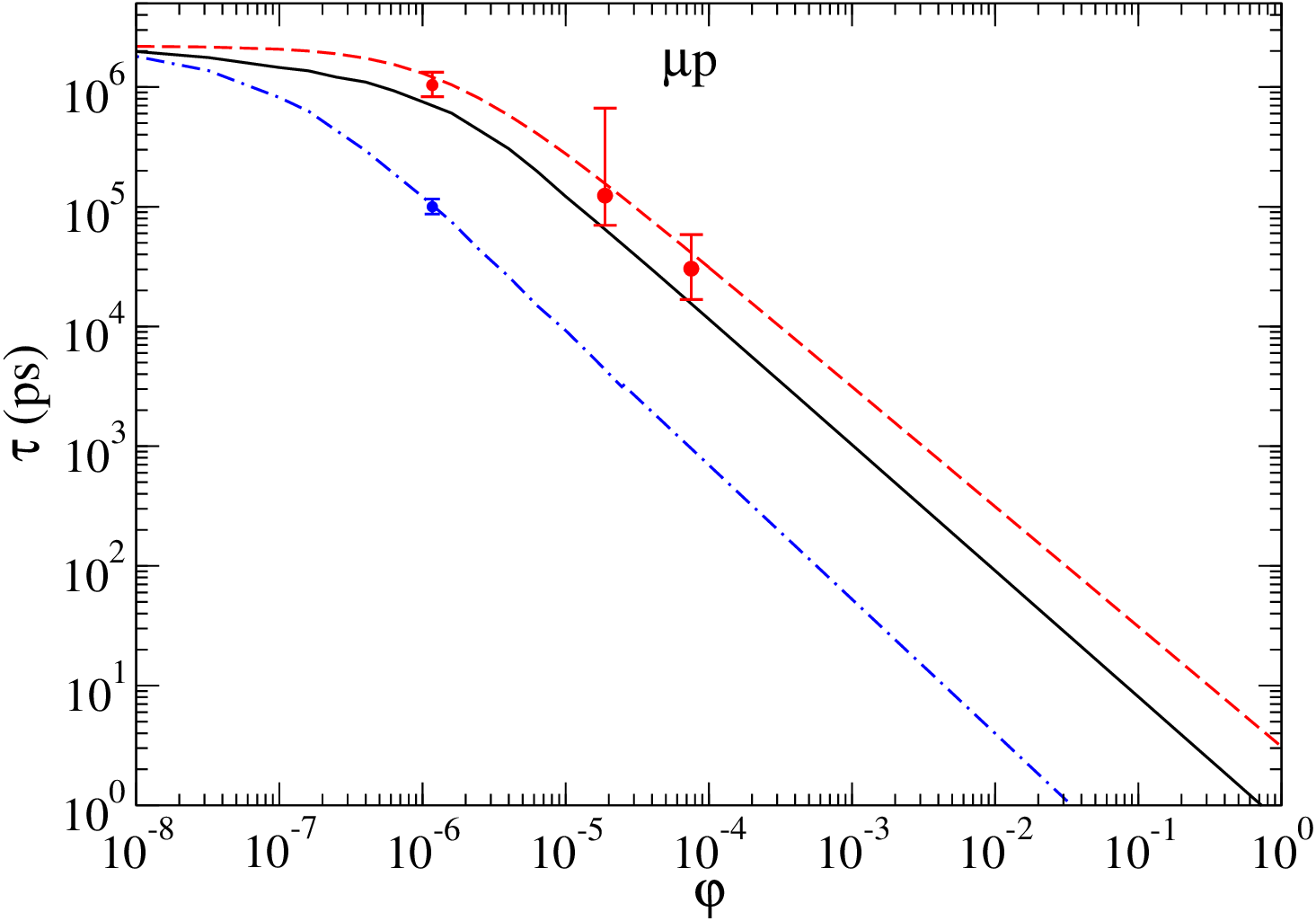}}
\caption{  The density dependence of the $(\mu^- p)_{2s}$ 
lifetime calculated for various fractions: below the $2s-2p$ threshold - dashed line,
above the $2s-2p$ threshold - dashed-dotted line. The lifetime, $\tau_{2s}$, 
of the $2s$ state independently whether the 
kinetic energy of muonic atom is above or below the $2s-2p$ threshold 
is shown by the solid line. The experimental values from \cite {PDH}}.
\label{lifetimemup}
\end{figure} 

\begin{figure}[!ht]          
\centerline{
\includegraphics[width=\columnwidth,keepaspectratio]
{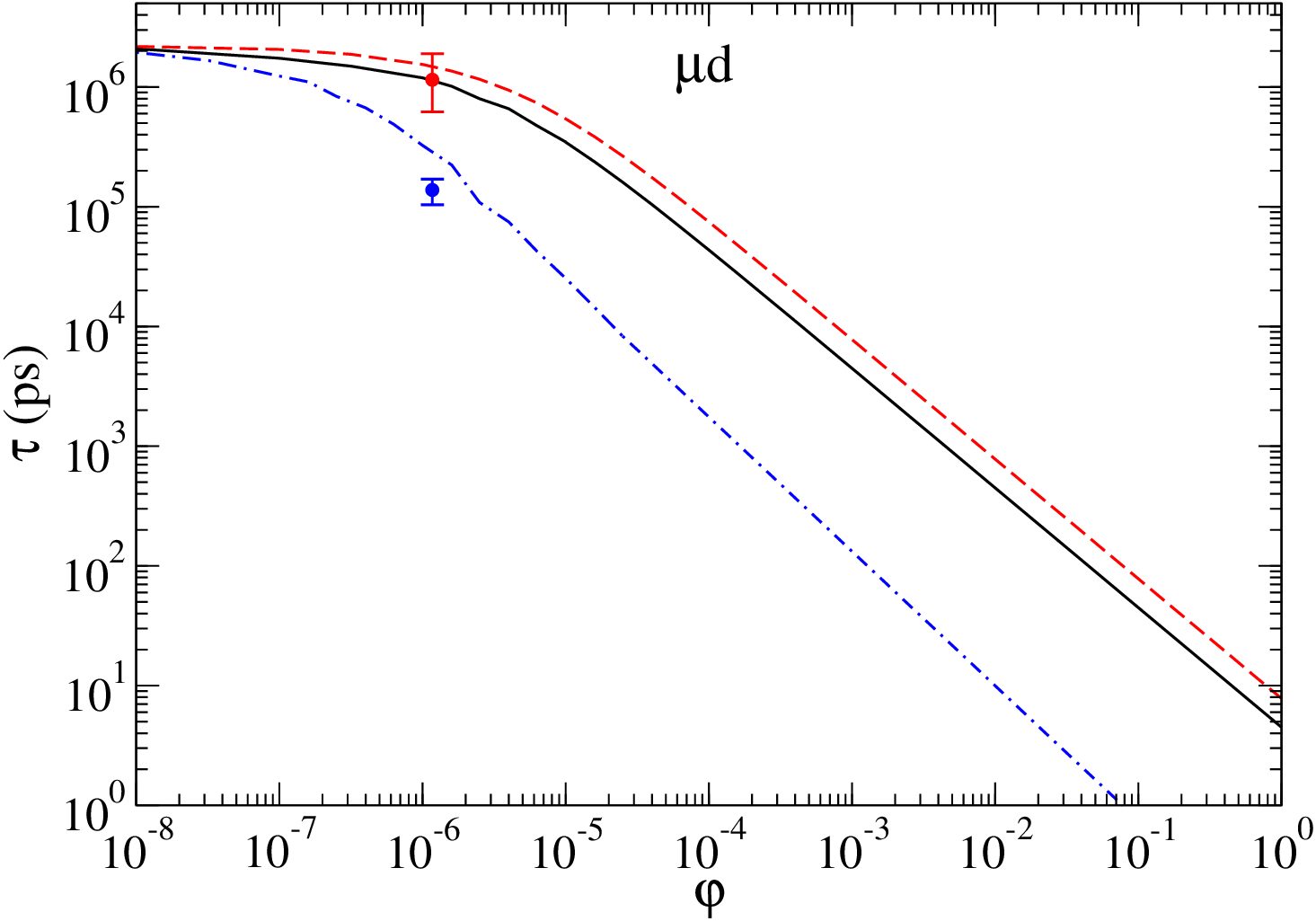}}
\caption{  The same as in Fig.~\ref{lifetimemup}
for $(\mu^- d)_{2s}$. The experimental values from \cite{Mud}}.
\label{lifetimemud}
\end{figure} 

Results of calculations are presented in Figs. \ref{lifetimemup} and \ref{lifetimemud} for the $(\mu^- p)_{2s}$ and $(\mu^- d)_{2s}$ atoms, respectively. As it is seen from the figures, the lifetime of the $2s$ state for both muonic hydrogen and muonic deuterium has a similar dependence on density for both fractions. At the lowest densities, the lifetimes are determined by the muon decay rate.
With increasing density, the role of collisional processes increases, which leads to a decrease in the lifetime of both fractions. The lifetime for the fraction with kinetic energy below the $2s-2p$ threshold, $\tau_{2s}^{\rm b}$, decreases much more slowly with increasing target density than for the fraction with kinetic energy above the $2s-2p$ threshold. At a densities  $\varphi>10^{-5}$,  the lifetime of a $2s$ state behaves as $\varphi^{-1}$,  and the ratio $\tau_{2s}^{\rm b}$/$\tau_{2s}^{\rm a}$  is  equal to about 15. 

In the case of muonic hydrogen, the theoretical values of the lifetime of the fractions with kinetic energy both above and below the $2s-2p$ threshold are in excellent agreement with the available experimental data. 
In the case of muonic deuterium, the agreement between the theoretical and available at one density experimental values of the lifetimes of the $2s$ state can be considered as good for $\tau_{2s}^{\rm b}$ and satisfactory for $\tau_{2s}^{\rm a}$.

\section{Conclusion}

The fully quantum-mechanical study of the collisional processes --- elastic scattering, collision-induced radiative quenching, and Coulomb deexcitation --- has been performed in the framework of the close-coupling approach for $(\mu^- p)_{2s}+H$ and $(\mu^- d)_{2s}+D$ collisions at kinetic energies below the $2s-2p$ threshold. 
The cross sections have been calculated with the basis set including all open and closed channels associated with the exotic atom states with the principal quantum number up to $n_{max}=30$ and then have been used as input data in the kinetics of the atomic cascade.

An important result of this work is the elucidation of the role of the collision-induced radiative quenching in the destruction of a muonic atom in $2s$-state. As far as the authors know, this problem has not been previously investigated.
Our calculations showed that this process makes a significant contribution to the total disintegration of the $2s$-state of muonic atoms at energies below the  $2s-2p$ threshold. 

The kinetics of the atomic cascade has been investigated for  both muonic hydrogen and muonic deuterium atoms within the updated version of the cascade model, which has a number of essential improvements over the previous cascade calculations.
The cascade calculations have been performed for target
densities covering eight orders of magnitude from $10^{-8}$ up to 1 (in the units of the liquid hydrogen density).

As a result of {\it ab  initio}  cascade calculations, a number of characteristics of the $2s$ state of muonic hydrogen and deuterium atoms were obtained: primary populations and kinetic energy
 distributions, absolute values of intensities different modes of disintegration, the probability of decay of the $2s$ state through different channels and lifetimes of the $2s$ state at kinetic energy both above and below the $2s-2p$ threshold. The obtained theoretical results are compared with the available experimental data. The results of the work can be used for the planning experiments and analysis of experimental data.

\end{document}